\documentclass[12pt]{article}
\usepackage{amsmath,amssymb,amsthm,amscd,dsfont,graphics,color}
\input epsf.sty
\newcommand{\CC}{{\cal C}}

\newcommand{\CK}{{\cal K}}
\newcommand{\CL}{{\cal L}}

\newcommand{\CN}{{\cal N}}
\newcommand{\CO}{{\cal O}}

\theoremstyle{definition}

\def\IZ{{\mathbb Z}}

\def\IC{{\mathbb C}}

\newcommand{\tr}{{\rm Tr}}
\newcommand{\re}{{\rm e}}
\newcommand{\ri}{{\rm i}}
\newcommand{\rd}{{\rm d}}
\newcommand{\Dd}{\Delta_{21}}

\newcommand{\be}{\begin{equation}}
\newcommand{\ee}{\end{equation}}
\newcommand{\ba}{\begin{aligned}}
\newcommand{\ea}{\end{aligned}}
\newcommand{\bea}{\begin{eqnarray}\displaystyle}
\newcommand{\eea}{\end{eqnarray}}

\newcommand{\p}{\partial}
\newcommand{\sectiono}[1]{\section{#1}\setcounter{equation}{0}}

\newdimen\tableauside\tableauside=1.0ex
\newdimen\tableaurule\tableaurule=0.4pt
\newdimen\tableaustep
\def\phantomhrule#1{\hbox{\vbox to0pt{\hrule height\tableaurule width#1\vss}}}
\def\phantomvrule#1{\vbox{\hbox to0pt{\vrule width\tableaurule height#1\hss}}}
\def\sqr{\vbox{%
  \phantomhrule\tableaustep
  \hbox{\phantomvrule\tableaustep\kern\tableaustep\phantomvrule\tableaustep}%
  \hbox{\vbox{\phantomhrule\tableauside}\kern-\tableaurule}}}
\def\squares#1{\hbox{\count0=#1\noindent\loop\sqr
  \advance\count0 by-1 \ifnum\count0>0\repeat}}
\def\tableau#1{\vcenter{\offinterlineskip
  \tableaustep=\tableauside\advance\tableaustep by-\tableaurule
  \kern\normallineskip\hbox
    {\kern\normallineskip\vbox
      {\gettableau#1 0 }%
     \kern\normallineskip\kern\tableaurule}%
  \kern\normallineskip\kern\tableaurule}}
\def\gettableau#1{\ifnum#1=0\let\next=\null\else
\squares{#1}\let\next=\gettableau\fi\next}

\newcommand{\figref}[1]{Fig.~\protect\ref{#1}}
\begin{document}

%%%%%%%%%%%%%%%%%%%%%%%%%%%%%%%%%%%%%%%%%%%%%%%%%%%%%%%%%%%%%%%%%%%%%%%%%%%%%%%%%

\baselineskip=18pt

\begin{titlepage}
\begin{flushright}
\parbox[t]{1.8in}{
BONN-TH-2010-01
}
\end{flushright}

\begin{center}

\vspace*{ 1.2cm}

{\bf \Large Direct Integration and Non-Perturbative Effects\\[.3cm] in Matrix Models}

\vskip 1.2cm

\begin{center}
\bf{Albrecht Klemm\textsuperscript{{\em a},}\footnote{~\texttt{aklemm@th.physik.uni-bonn.de}}, 
Marcos Mari\~no\textsuperscript{{\em b},}\footnote{~\texttt{marcos.marino@unige.ch}} 
and Marco Rauch\textsuperscript{{\em a},}\footnote{~\texttt{rauch@th.physik.uni-bonn.de}}}
\end{center}
\vskip 0.2cm

{\em \textsuperscript{a}Bethe Center for Theoretical Physics, \\
Universit\"at Bonn, Nu\ss allee 12, D-53115 Bonn, Germany}\\[.3cm]
{\em \textsuperscript{b}D\'epartement de Physique Th\'eorique et Section de Math\'ematiques, \\
Universit\'e de Gen\`eve, CH-1211 Gen\`eve, Switzerland}

\vspace*{.1cm}

\end{center}

\vskip 0.2cm

\begin{center} {\bf Abstract } \end{center}

We show how direct integration can be used to solve the closed amplitudes 
of multi-cut matrix models with polynomial potentials. In the case of the cubic matrix model, 
we give explicit expressions for the ring of non-holomorphic modular objects that are needed 
to express all closed matrix model amplitudes. This 
allows us to integrate the holomorphic anomaly equation up to holomorphic 
modular terms that we fix by the gap condition up to genus four. 
There is an one-dimensional submanifold of the moduli space in which the spectral curve 
becomes the Seiberg--Witten curve and the ring reduces to the non-holomorphic 
modular ring of the group $\Gamma(2)$. On that submanifold, the gap 
conditions completely fix the holomorphic ambiguity and the model can be solved explicitly 
to very high genus. We use these results to make precision tests of the connection between the large order 
behavior of the $1/N$ expansion and non-perturbative effects due to instantons. 
Finally, we argue that a full understanding of the large genus asymptotics in the multi-cut case 
requires a new class of non-perturbative sectors in the matrix model.

\end{titlepage}

%%%%%%%%%%%%%%%%%%%%%%%%%%%%%%%%%%%%%%%%%%%%%%%%%%%%%%%%%%%%%%%%%%%%%%%%%%%%%%%%%

\tableofcontents
\newpage

%%%%%%%%%%%%%%%%%%%%%%%%%%%%%%%%%%%%%%%%%%%%%%%%%%%%%%%%%%%%%%%%%%%%%%%%%%%%%%%%%

\sectiono{Introduction and Results}
\label{sec:intro}

In this paper we propose direct integration as a new method to solve the closed 
amplitudes for multi-cut matrix models with polynomial potentials. More precisely, 
we calculate the closed partition function of such matrix models
\be
Z(\underline{S})=\exp\left(\sum_{g} g_s^{2g-2} F_g(\underline{S})\right)
\ee
perturbatively in the genus $g$, but exactly in the 't Hooft 
parameters $\underline{S}$. Exact means that the $F_g(\underline{S})$ 
are given in terms of period integrals of the spectral curve and 
can be written explicitly in terms of modular forms of subgroups 
of SP$(2 g, \mathbb{Z})$.    
    
Direct integration refers to a method of solving the holomorphic anomaly 
equation~\cite{bcov} using the modular transformation properties of the 
amplitudes under the monodromy group of the spectral curve. This method has been developed in the 
context of topological string theory in \cite{Yamaguchi,hkq,gkmw,al,Hag,hkK3}. The fact that the 
holomorphic anomaly equations govern such matrix models was suggested by the large 
$N$ duality of \cite{dv}. In this duality, type B topological string amplitudes on 
certain local Calabi-Yau spaces turn out to be encoded in the $1/N$ expansion of matrix 
model partition functions. Therefore, the holomorphic anomaly of the topological 
string naturally carries over to these matrix models as first pointed out in ~\cite{hk}. It has been 
shown much more generally in~\cite{emo} that the holomorphic anomaly equation 
is valid for all matrix models which are solvable by the method of~\cite{eo}.  

The holomorphic anomaly equation relates anti-holomorphic derivatives of the closed amplitudes 
$F_g(\underline{S})$ at genus $g$ to lower genus amplitudes 
$F_{h<g}(\underline{S})$, in a recursive way. Since only the anti-holomorphic derivative is specified 
by the equations, the procedure leaves a holomorphic ambiguity, i.e.~$F_g(\underline{S})
=F^{\rm nh}_g(\underline{S})+f_g(\underline{S})$ splits into a non-holomorphic term 
$F^{\rm nh}_g(\underline{S})$, which is determined by the holomorphic anomaly equation, 
and the holomorphic ambiguity $f_g(\underline{S})$, which must be fixed genus by genus 
by using modular properties and boundary conditions 
at special points in the moduli space. The modular 
transformation properties imply that the amplitudes are generated by a finite ring of modular forms, 
which have holomorphic as well as 
non-holomorphic generators. Modularity and the holomorphic anomaly equation imply that 
the total amplitude $F_g(\underline{S})$ is a polynomial in these generators whose degree grows linearly 
with the genus. The ambiguity  $f_g(\underline{S})$ is a polynomial generated by the 
smaller ring of holomorphic generators. The finite number of coefficients in 
this polynomial must be fixed by boundary conditions. 

In this paper we find that the gap conditions, 
which where investigated in non-compact~\cite{hk,hkr,alm,hk2} and compact Calabi-Yau 
backgrounds~\cite{hkq,gkmw,Hag,hkK3}, provide enough independent 
boundary conditions to fix the ambiguity (and hence the amplitudes) 
completely. Following \cite{hkr} we refer to this property as 
integrability of the holomorphic anomaly equation.
    
The large $N$ duality relating matrix models and topological strings gives a natural geometric interpretation 
to the algebraic objects describing the planar limit of the matrix model \cite{dv}. The spectral curve $y(x)$ of the matrix model 
(which, in the case of polynomial potentials, is a hyperelliptic curve) describes the distribution of eigenvalues in the planar limit, 
and in the topological string dual it describes the nontrivial part of the Calabi--Yau geometry. We derive the modular ring starting from the Picard-Fuchs equations governing the periods of the 
form $\Omega=y(x)\rd x$. This is a general method\footnote{For example, a meromorphic modular form 
of weight $k$ of SL$(2,\mathbb{Z})$ or a congruence subgroup fulfills a linear 
differential equation of order $k+1$ in the total modular invariant~\cite{Zagier}.}, and 
since we expect that the gap boundary conditions fix the ambiguity, our approach should 
apply to general multi-cut matrix models with polynomial potential.  

Of course, the formalism of \cite{eo} gives in principle all the genus $g$ free energies of generic multi-cut matrix model in terms of universal 
formulae on the spectral curve. The price to pay for such a general approach is that its detailed implementation is 
in practice very involved. Even in two-cut models, going beyond genus two with the methods of \cite{eo} is not very feasible. In contrast, direct integration 
becomes very powerful when the spectral curve and its modular group are simple. 

In this paper, in order to illustrate the method of direct integration, we focus on the two-cut  
matrix model with a cubic potential. In this model the $N$ eigenvalues split in two 
sets $N=N_1+N_2$ and condense in sets near the two critical points of the 
potential. This leads to the cuts in the spectral curve shown in figure~\ref{FigCycles}. 
There are two independent 't Hooft couplings $S_i=g_s N_i$, $i=1,2$, 
which correspond to the integrals of $\Omega$ over the two cuts. As shown in \cite{dv}, the planar free energy of this matrix model, 
$F_0(S_1, S_2)$, calculates 
the exact superpotential $W_{\rm eff}$ of an $\CN=2$ $U(M)$ supersymmetric gauge theory broken down to an $\CN=1$ gauge 
theory $U(M_1)\times U(M_2)$, by a cubic three-level superpotential 
in the adjoint~\cite{dv} (notice that $N_i$ are unrelated to $M_i$). The higher genus amplitudes $F_g(S_1, S_2)$ in the 
matrix model arise as generalized couplings in a non-commutative deformation of the $\CN=1$ gauge theory~\cite{ov-ncgrav}.

Certain aspects of the original $\CN=2$ theory can be recovered from the $\CN=1$ 
theory by breaking the gauge symmetry to the Cartan subgroup and taking the limit 
in which the superpotential vanishes \cite{cv}. When the gauge group is $SU(2)$, 
a cubic superpotential is enough to go to the Coulomb branch. This implies that various quantities appearing in the Seiberg--Witten solution 
of pure $\CN=2$ super Yang--Mills theory \cite{sw} can be obtained from a matrix model calculation with a cubic potential, and on the slice $S_1=-S_2$. These include the gauge coupling~\cite{dgkv} and the $R_+^2$ gravitational coupling \cite{kmt,dst}. In fact, the spectral curve of the cubic matrix model on that slice is identical to the Seiberg--Witten curve \cite{dgkv}. Since the modular group of this curve is particularly simple, direct integration becomes 
an extremely powerful method to calculate the $F_g(S_1,-S_1)$, as we show in section \ref{sec:slice}. 
  
On the other hand, in the $SU(2)$, $\CN=2$ gauge theory there is an infinite number of couplings $F_g(a)$, $g\ge 2$, which describe the gauge-gravity couplings $F_+^{2g-2} R_+^2$ 
involving the graviphoton field strength $F_+$. These couplings appear naturally in Nekrasov's partition function \cite{nekrasov} and they can be 
also obtained by using the holomorphic anomaly equations. This was shown for the pure gauge theory and $SU(2)$ with matter in~\cite{hk,gkmw} and~\cite{hk2} respectively. However, it was noticed in \cite{kmt} that these higher genus couplings $F_g(a)$ 
do not agree with the higher genus $F_g(S_1, S_2)$ obtained in the cubic matrix model and then restricted to the slice $S_2=-S_1$. 
This disagreement is due to the fact that the Seiberg-Witten differential $\lambda_{\rm SW}$ differs from the natural differential $\Omega$ on the spectral 
curve of the matrix model. In contrast, $\tau$ and $F_1$ only depend on the spectral curve, and not on the differential, and therefore are the same in both cases. In \cite{KlemmSulk,Sulkowski} matrix models are derived which encode all $\CN=2$ gauge theory amplitudes $F_g$ for arbitrary $g$, however one has to introduce potentials involving polylogarithms and their quantum generalizations.

An interesting application of our computation of the couplings $F_g(S_1, S_2)$ at high genus is the study of non-perturbative effects in matrix models and their 
connection to the large order behavior of the $1/N$ expansion. It is well-known that, in many quantum systems, 
there is a connection between perturbation theory at large orders 
and instantons (see for example \cite{zj}). In matrix models, instanton configurations correspond to the tunneling of eigenvalues between different 
saddle points \cite{david,shenker}. A detailed analysis of these configurations for off-critical, one-cut matrix models can be found in \cite{mswone}, which verified the connection to the large 
order behavior of the $1/N$ expansion in detail in some nontrivial examples. In this paper we explore this connection in the two-cut matrix model. On the one hand, we find that the 
large order behavior is controlled at leading order by the action of a single eigenvalue tunneling from one saddle-point to the other, in agreement with the general ideas put forward in 
\cite{david,shenker,mswone}. On the the other hand, we argue that a full understanding of this connection requires new non-perturbative sectors which have not been yet identified in the 
matrix model. The existence of these sectors is also suggested by a recent analysis of the asymptotic behavior of the instanton solutions of the Painlev\'e I equation \cite{gikm}. We conjecture that these sectors might involve topological brane-antibrane systems. 

The paper is organized as follows. In section~\ref{sec:holanddirectint} we 
give the general ideas of the direct integration method and of the modular covariant approach of \cite{abk}. 
In section 3 we describe in detail the geometry underlying the two-cut cubic matrix model, we set up the direct 
integration formalism and we analyze the boundary 
conditions. In section 4 we point out that the modular covariant formulation is 
most powerful on the one-dimensional slice of moduli space $S_1=-S_2$, and we develop direct integration on this 
submanifold in terms of non-holomorphic modular forms of $\Gamma(2)$. Section 5 is devoted to the analysis of 
instanton effects in the two-cubic matrix model, and their connection to the large order of the $1/N$ expansion. 
Finally, the Appendices collect some useful information about the cubic matrix model as well as results on modular forms which are 
used in the paper.

%%%%%%%%%%%%%%%%%%%%%%%%%%%%%%%%%%%%%%%%%%%%%%%%%%%%%%%%%%%%%%%%%%%%%%%%%%%%
\sectiono{Holomorphic anomaly and direct integration}
\label{sec:holanddirectint}

Below we review very briefly the generic aspects of the techniques of direct integration of the holomorphic 
anomaly equation of~\cite{bcov}
\be\label{HAEq}
\bar{\partial}_{\bar\imath} F_g=\frac{1}{2}\bar{C}_{\bar\imath}^{jk}\left(D_jD_kF_{g-1}+
\sum_{r=1}^{g-1}D_jF_{g-r} D_kF_r\right),\qquad (g>1)\ ,
\ee
which was derived for Calabi-Yau three-folds in \cite{Yamaguchi,hkq,gkmw,al,Hag,hkK3}. In our application to the spectral curve $\Sigma$ of a matrix model, the 
$D_i$ are covariant derivatives $D_i$ with respect to the metric $G$ on the moduli space of the Riemann surface $\Sigma$. 

We note that $\bar{C}_{\bar\imath}^{jk}=\bar C_{\bar \imath \jmath \bar k} G^{\bar \jmath j}G^{\bar kk}$, where $C_{ijk}$ can be derived from the holomorphic prepotential $F_0$ as $C_{ijk}=D_i D_j \partial_k F_0$. The prepotential $F_0$, the metric 
$G_{i\bar \jmath}$ and flat coordinates $\underline{S}$ can all be derived from the period integrals 
\be\label{symplectic}
\left(\int_{a^i} \Omega, \int_{b_i}\Omega\right), \qquad i=1,\ldots,g(\Sigma)
\ee
over a symplectic basis $(a^i,b_i)$ of $H_1(\Sigma,\mathbb{Z})$. In particular, given a point in the moduli space, one 
can make a choice of this symplectic basis, so that suitable flat coordinates are defined by
\be
S^i=\int_{a^i}\Omega
\label{coord}
\ee
while the $b_i$ periods $\Pi_i$ fulfill   
\be 
\Pi_i=\frac{\partial F_0}{\partial S^i} \ .
\label{derivatives}
\ee
These relations determine the prepotential $F_0$ up to an 
irrelevant constant. We define the $\tau$ matrix of the Riemann 
surface as
\be\label{tauij}
\tau_{ij}=\frac{\partial^2F_0}{\partial S^i \partial S^j}.
\ee
The matrix ${\rm Im}(\tau)_{ij}$ is positive definite, and it gives the metric on the moduli space 
of the model. Equivalently, the metric can be obtained form 
the K\"ahler potential
\be
K=\frac{1}{2 \pi i}\left(\Pi_i {\bar S}^{\bar \imath} - {\bar \Pi}_{\bar \imath} S^i \right).
\ee

On Riemann surfaces the period integrals can often be directly  performed. Alternatively it might be useful to derive the Picard-Fuchs equations and reconstruct the periods as linear combinations of their solutions. Much of the above has been spelled out 
in the context of the Riemann surfaces for the B-model of topological string theory on non-compact Calabi-Yau in~\cite{hkr}. The relevant compact part of the geometry is given by a Riemann surface and a meromorphic differential, which comes from reducing 
the holomorphic $(3,0)$-form on the Riemann surface. After identification of the former with the spectral curve 
$\Sigma$ and the later with the form $\Omega$, we can use the formalism discussed in~\cite{hkr}. 

One property of the matrix model geometry is that the periods over the $a$-cycles do not fulfill the relation $\sum_{i=1}^r S^i=0$. 
Usually this relation is inherited by the periods of holomorphic forms due to the homological relation of the cycles. 
However, in matrix models one has $\sum_{i=1}^r S^i\propto N$, because $\Omega$ has one non-vanishing residue outside the cuts. 
This leads to one algebraic relation between the periods in terms of the $r$ parameters, which for the 
$r=2$ case (the cubic matrix model) is expressed in eq.~(\ref{t2zexact}). The property of a non-vanishing residue is shared with Seiberg-Witten 
theories with matter~\cite{hk2} and certain non-compact Calabi-Yau geometries with more than one K\"ahler class~\cite{hkr,alm}.

\subsection{Direct integration}
%-----------------------------------
\label{sec:directint}
The so-called propagator plays a decisive role in the solutions of the B-model \cite{bcov}. For the formalism on the Riemann surface $\Sigma$ one needs only one type\footnote{In the threefold cases there are three 
types $S^{ij},S^i$ and $S$.} of propagator $S^{ij}$ defined by
\be 
\label{def:propagator}
{\bar \partial}_{\bar \imath} S^{ij}=C_{\bar \imath}^{ij},     
\ee
where $i,j=1,\ldots, r$ and  $r$ is the number of parameters in the model.
Following~\cite{bcov} it can be shown that the $F_g$ can be written as
\be
\label{generalform}
F_g=\sum_{|I|=0}^{3g-3} f_{g,i_1 \ldots i_{|I|}}(\underline{S})\ S^{i_1 i_2}\ldots S^{i_{|I|-1} i_{|I|}}\ 
\ee 
where the $f_{g,I}(\underline{S})$ are holomorphic tensors of the moduli.  
The most important property of the $S^{ij}$ is that
\be 
\partial_{\bar \imath} F_g=C_{\bar \imath}^{ij} \frac{\partial F_g}{\partial S^{ij}}.  
\ee
If one assumes linear independence of the $S^{ij}$ as functions of $\underline{S}$, it follows from this property that (\ref{HAEq}) can be rewritten as a set of equations
\be\label{HAEq2}
\frac{\partial F_g}{\partial S^{ij}} =\frac{1}{2}\left(D_jD_kF_{g-1}+\sum_{r=1}^{g-1}D_jF_{g-r} 
D_kF_r\right),\qquad (g>1)\ .
\ee
These equations can be integrated algebraically, provided that the r.h.s. can be expressed in 
terms of the $S^{ij}$ contracted  by holomorphic  tensors as in the r.h.s of (\ref{generalform}).
This is possible since the following closing relations are fulfilled due to special geometry \cite{bcov,al}
\begin{equation} \label{DS}
D_i S^{kl} = -C_{inm} S^{km} S^{ln}+ f^{kl}_i ,
\end{equation}
\begin{equation} \label{PropEq}
\Gamma^k_{ij}=- C_{ijl} S^{kl}+\tilde f_{ij}^k\ ,
\end{equation}
\begin{equation}\label{F1prop}
\partial_i F_1=\frac{1}{2} C_{ijk}S^{jk}+A_i.
\end{equation}
Here the $f^{kl}_i$, $\tilde f^{kl}_i$ and $A_i$ are holomorphic ambiguities, which must have the same transformation 
properties  as the expressions on the left-hand side. These ambiguities are due to the fact that (\ref{def:propagator}) 
defines $S^{ij}$ only up to an holomorphic tensor. Different 
choices are possible and lead to a redefinition of the $f_{g,I}$ 
in (\ref{generalform}). As we mentioned above the periods are not algebraically independent, see for example (\ref{t2zexact}). As a 
consequence it is possible to make a choice for the above 
ambiguities so that for a given $i$ one has $S^{ik}=0$, 
$\forall k$, i.e. the matrix of propagators has 
effectively only rank $\rho=r-1$. We call the auxiliary parameter $t$ . There may be more auxiliary parameters stemming from the independent non-vanishing residua of $\Omega$. If there are $\kappa$ such residua, the rank is reduced to $\rho = r - \kappa$.

Whether one works with the redundant or the reduced set of 
propagators the  equation (\ref{HAEq2}) can easily be integrated w.r.t. $S^{ij}$ and $F_g$ becomes of degree $3g-3$ in the 
$S^{ij}$. This is an efficient way to solve the recursion, but 
at each step one still has to determine the holomorphic ambiguity. 

\subsection{Modular covariant formulation}
%------------------------------------------
\label{sec:modularcovariant} 
It is possible to relate the non-redundant set of propagators 
to quasimodular forms. In particular, in the holomorphic polarisation, 
the following properties derived in~\cite{abk} hold:

\begin{enumerate}
\item $F_g(\underline{S})$ is invariant under the monodromy 
group $\Gamma$ of the Riemann surface $\Sigma$.     
\item $F_g(\underline{S})$ is an {\sl almost}-holomorphic modular function, i.e. its non-holomorphic dependence is encoded solely 
in $((\tau -\bar \tau)^{-1})^{IJ}$, where $I,J=1,\ldots,\rho$ and  $\tau_{IJ}$ is the standard matrix valued 
modular parameter living in the Siegel upper half space.\footnote{$\rho = r - \kappa$, where $\kappa$ is the number  of
  independent non-vanishing  residua of $\Omega$.}
\item The non-holomorphic dependence combines always with quasimodular forms $E^{IJ}$ to give almost-holomorphic modular forms
\be
{\widehat E}^{IJ}=E^{IJ}(\tau)+((\tau -\bar \tau)^{-1})^{IJ}.
\ee
Here we defined $E^{IJ}(\tau)$ as derivative of ${\partial \over \partial \tau_{IJ}} F_1(\tau)$. The anomaly 
equation of~\cite{bcov} for 
$F_1(\tau)$ implies that this is a non-holomorphic modular 
invariant 
\begin{equation}
F_1=-\log\left[{\rm \det}^{1\over 2}({\rm Im}(\tau_{IJ})) \left( \bar \Phi_k({\bar \tau})\Phi_k(\tau)\right)^a\right]
\end{equation} 
under the monodromy group $\Gamma$. $\Phi_k(\tau)$ is a holomorphic Siegel modular cusp form of weight $k$ 
which vanishes at the discriminant $\Delta$ of the Riemann surface. It transforms as 
$\Phi_k(\tau_\gamma)=\det(C\tau+D)^k \Phi_k(\tau)$,  where $\tau_\gamma$ is given by
\be
\tau_\gamma=\left(A\tau + B\right) 
\left(C \tau + D\right)^{-1},\quad \gamma=\left(\begin{array}{cc} A & B\\ C & D \end{array}\right)\in {\rm Sp}(2 \rho, \mathbb{Z}) \ .
\ee
Such modular forms exist for all genus and can be written as products of even theta functions~\cite{klingen}.   
The exponent $a$ will make the argument of the log invariant and the vanishing order at the discriminant 
$\frac{1}{12}\log(\Delta)$. For an elliptic curve $\Phi_k(\tau)$ is typically the Dedekind $\eta$-function. 
However, if the subgroup $\Gamma$ allows for several cusp forms, $\Phi_k(\tau)$ can be a suitable multiplicative combination
of them. In virtue of the definition  ${\widehat E}^{IJ}$ transforms as a Siegel modular form
\be 
{\widehat E}^{IJ}(\tau_\gamma)=(C \tau + D)^I_K (C\tau +D)^{J}_{L} {\widehat E}^{KL}(\tau)\ .
\ee
%where
%\be 
%\gamma=\left(\begin{array}{cc} A & B\\ C & D \end{array}\right)\in {\rm Sp}(2 \rho, \mathbb{Z})\ .
%\ee
%Here we used $\rho=r-1$ and the action of the monodromy group on $\tau$ is given by the generalized M\"obius transformation
%\be 
%\tau\rightarrow \tau_\gamma=\left(A\tau + B\right) 
%\left(C \tau + D\right)^{-1} \ .
%\ee 
\item $F_g(\underline{S})$ can be expanded as 
\be 
F_g=\sum_{|I|=0}^{3g-3} \tilde f_{g,I_1,\ldots,I_{|I|}} {\widehat E}^{I_1 I_2}\ldots {\widehat E}^{I_{|I|-1} I_{|I|}}\  . 
\ee
\end{enumerate}
Note that $\tilde f_{g,I}$ has to compensate for the modular transformation of $\tau$ and can in principle be expressed through
holomorphic modular forms.
%However $f_{g,I}$ through $\tau$ and explicetly on the auxiliary parameter and we find simple expressions only when $s=0$.      
 
%%%%%%%%%%%%%%%%%%%%%%%%%%%%%%%%%%%%%%%%%%%%%%%%%%%%%%%%%%%%%%%%%%%%%%%%%%%%
\sectiono{The two-cut cubic matrix model}
\label{sec:2cutcubicMM}

As shown by Dijkgraaf and Vafa in \cite{dv}, the B-model topological string theory on certain non-compact Calabi--Yau geometries is captured by a matrix model. The matrix model is the $n$-cut matrix model with potential $W(x)$, while the Calabi--Yau geometry is the following hypersurface in $\IC^4$
\be\label{eq:localCYgeo}
uv=y^2-(W'(x)^2+f(x)).
\ee
Here, $f(x)$ is a polynomial of degree $n-1$ that splits the $n$ double zeroes of $W'(x)^2$, see \cite{mmhouches} for a detailed review. In the following we will combine the 
Dijkgraaf--Vafa correspondence with known results about the holomorphic anomaly equation in order to give a recursive solution of multi-cut matrix models.

\subsection{The geometrical setup}
%------------------------------------------------
\label{sec:geometricalsetup} 
In this paper we will be interested in multi-cut, Hermitian matrix models. The partition function is defined by
\be
\label{zmi}
Z={1\over {\rm vol}(U(N))} \int \rd M \, \re^{-{1\over g_s} \tr W(M)} 
\ee
where $W(M)$ is a polynomial of degree $d=n+1$ in the 
$N\times N$ matrix $M$. The most general saddle point of this model at large $N$ is a multi-cut solution, in 
which the eigenvalues of $M$ condense along cuts 
\be
[a_i^{-}, a_i^+]\subset \IC, \qquad i=1, \cdots, n,
\ee
in the complex plane. The cuts are centered around the $n$ critical points of $W(x)$. One way 
of encoding the planar solution of the matrix model is through its resolvent
\be
\omega(x) ={1\over N} \Big\langle \tr {1\over x-M}\Big\rangle.
\ee
The planar limit of this correlator, denoted by $\omega_0(x)$, has the structure (see for example \cite{dfgzj})
\be
\omega_0(x)={1\over 2t} (W'(x)-y(x)),
\ee
where 
\be
t=g_s N
\ee
is the total 't Hooft parameter, and 
\be
\label{sc}
y^2(x)=\left(W'(x)\right)^2 + f(x) =c \prod_{i=1}^n (x-a_i^-) (x-a_i^+)
\ee
is called the {\it spectral curve} of the multi-cut matrix model. In the matrix model literature it is customary to write it as
\be
\label{scsigma}
y^2(x)=M(x) {\sqrt{\sigma(x)}}, 
\ee
where $\sigma(x)$ is a polynomial in $x$,
\be
\sigma(x)=\prod_{i=1}^{2s}(x-x_i),
\ee
and $s\le n$. Of course, if all the roots in (\ref{sc}) are different, $s=n$ and $M(x)$ is a constant. 

Through the large $N$ duality of \cite{dv}, the spectral curve (\ref{sc}) describes 
the nontrivial part of the target geometry (\ref{eq:localCYgeo}). The positions of the endpoints are fixed by the asymptotic condition 
\be
\omega_0(x) \sim {1\over x}, \qquad x \rightarrow \infty, 
\ee
and by the requirement that there are $N_i$ eigenvalues in each cut, 
\be
\label{filling}
{N_i \over N}={1\over 2}  \oint_{\CC_i}{\rd x \over 2\pi \ri}  \omega_0(x). 
\ee
In this equation, $\CC_i$ is a contour encircling the cut $[a_i^-, a_i^+]$ counterclockwise. The {\it partial 't Hooft parameters} $S_i$ are defined by
\be
S_i=g_s N_i. 
\ee
Notice that
\be
t=\sum_{i=1}^n S_i. 
\ee

\begin{figure}
\centering
\begin{picture}(0,0)%
\includegraphics{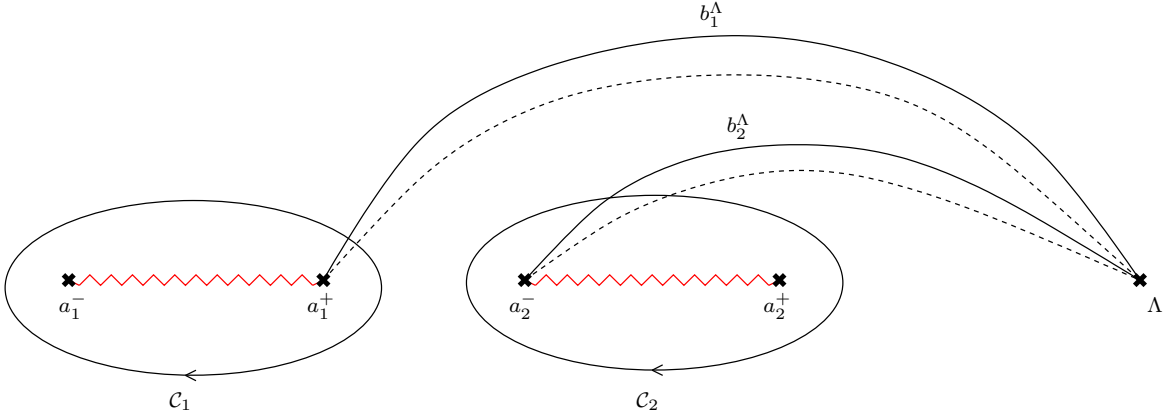}%
\end{picture}%
\setlength{\unitlength}{4144sp}%
\begingroup\makeatletter\ifx\SetFigFont\undefined%
\gdef\SetFigFont#1#2#3#4#5{%
  \reset@font\fontsize{#1}{#2pt}%
  \fontfamily{#3}\fontseries{#4}\fontshape{#5}%
  \selectfont}%
\fi\endgroup%
\begin{picture}(6862,2486)(5819,-4161)
\put(12658,-3537){\makebox(0,0)[lb]{\smash{{\SetFigFont{8}{9.6}{\rmdefault}{\mddefault}{\updefault}{\color[rgb]{0,0,0}$\Lambda$}%
}}}}
\put(8841,-3537){\makebox(0,0)[lb]{\smash{{\SetFigFont{8}{9.6}{\rmdefault}{\mddefault}{\updefault}{\color[rgb]{0,0,0}$a_2^-$}%
}}}}
\put(10364,-3537){\makebox(0,0)[lb]{\smash{{\SetFigFont{8}{9.6}{\rmdefault}{\mddefault}{\updefault}{\color[rgb]{0,0,0}$a_2^+$}%
}}}}
\put(7635,-3537){\makebox(0,0)[lb]{\smash{{\SetFigFont{8}{9.6}{\rmdefault}{\mddefault}{\updefault}{\color[rgb]{0,0,0}$a_1^+$}%
}}}}
\put(6149,-3537){\makebox(0,0)[lb]{\smash{{\SetFigFont{8}{9.6}{\rmdefault}{\mddefault}{\updefault}{\color[rgb]{0,0,0}$a_1^-$}%
}}}}
\put(9983,-1792){\makebox(0,0)[lb]{\smash{{\SetFigFont{8}{9.6}{\rmdefault}{\mddefault}{\updefault}{\color[rgb]{0,0,0}$b_1^\Lambda$}%
}}}}
\put(10142,-2458){\makebox(0,0)[lb]{\smash{{\SetFigFont{8}{9.6}{\rmdefault}{\mddefault}{\updefault}{\color[rgb]{0,0,0}$b_2^\Lambda$}%
}}}}
\put(9603,-4108){\makebox(0,0)[lb]{\smash{{\SetFigFont{8}{9.6}{\rmdefault}{\mddefault}{\updefault}{\color[rgb]{0,0,0}$\CC_2$}%
}}}}
\put(6810,-4108){\makebox(0,0)[lb]{\smash{{\SetFigFont{8}{9.6}{\rmdefault}{\mddefault}{\updefault}{\color[rgb]{0,0,0}$\CC_1$}%
}}}}
\end{picture}%
\caption{Choice of branch cuts and cycles on the elliptic geometry (\ref{ycurve}).}
\label{FigCycles}
\end{figure}

In the following we consider a cubic matrix model with potential $W$ given by
\be
\label{cubicpot}
W(x)= \frac{m}{2}x^2 + \frac{g}{3} x^3.
\ee
Since this model has two critical points $x=a_1$, $x=a_2$, the generic saddle will be a two-cut matrix model. If we write the matrix integral (\ref{zmi}) in terms of eigenvalues, we have to distinguish two different 
sets $\{ \mu_i\}_{i=1,\cdots, N_1}$, $\{ \nu_j\}_{j=1,\cdots, N_2}$, which are expanded around $a_1$, $a_2$, respectively, and we obtain 
\be
\label{exmint}
Z={1 \over N_1! N_2!} \int \prod_{i=1}^{N_1}\rd \mu_i \prod_{j=1}^{N_2} \rd \nu_j \prod_{i<j} \left( \mu_i -\mu_j \right)^2 \left( \nu_i -\nu_j \right)^2 
\prod_{i,j}  \left( \mu_i -\nu_j \right)^2 \re^{-{1\over g_s}\left(  \sum_i W(\mu_i) +\sum_j W(\nu_j)\right)}. 
\ee
Since
\be
W'(x)=m x + g x^2 =gx \Bigl( x + \frac{m}{g}\Bigr)=g(x-a_1)(x-a_2),
\ee
$W'(x)^2$ has two double zeroes at $x=a_1$, $a_2$, that are split by the degree one polynomial 
\be
\label{splitf}
f(x)=\lambda x+\mu
\ee
into four roots $a_1^{\pm},a_2^{\pm}$. Hence, the curve for the geometry/matrix model is given by
\be
\label{ycurve}
y^2=W'(x)^2+f=g^2(x-a_1^{-})(x-a_1^{+})(x-a_2^{-})(x-a_2^{+}).
\ee
We choose the branch cuts to be along the intervals $(a_1^{-}, a_1^{+})$ and $(a_2^{-}, a_2^{+})$, cf.~Fig.~\ref{FigCycles}.
It follows from (\ref{filling}) that the 't Hooft parameters for this curve are the periods of the one-form 
\be
\Omega = y(x)\, \rd x
\ee
around the branch cuts. Following the notation of \cite{civ}, we have
\be
S_i=\frac{1}{2\pi \ri} \int_{a_i^{-}}^{a_i^{+}} \Omega, \qquad \Pi_i=\frac{1}{2\pi \ri} \int_{b_i^{\Lambda}} \Omega.
\ee
These 't Hooft parameters are functions of the couplings in the potential $m$, $g$, and of the variables $\lambda,\mu$. Equivalently, 
they are functions of the branch points $a_i^{\pm}$ of the quartic curve (\ref{ycurve}). 
It is convenient to define new variables given by
\be
\begin{split}
z_1& = \frac{1}{4}(x_2 - x_1)^2,  \quad z_2 = \frac{1}{4}(x_4-x_3)^2,\\
Q& = \frac{1}{2}(x_1+x_2+x_3+x_4) = -\frac{m}{g},\\
I^2& = \frac{1}{4}\left[(x_3+x_4)-(x_1+x_2)\right]^2 =\left(\frac{m}{g}\right)^2-2(z_1+z_2),
\end{split}
\ee
where we label the cuts more conveniently as
\be
(a_1^{-},a_1^{+},a_2^{-},a_2^{+})=(x_1,x_2,x_3,x_4) 
\ee
and we also have
\be
\sigma(x)=\prod_{i=1}^4 (x-x_i). 
\label{oursigma}
\ee
We will use this in order to expand all 
four periods in powers of $z_1$ and $z_2$. 
Notice that $z_i$ are coordinates that parameterize 
the complex structure deformations of the local Calabi--Yau geometry (\ref{eq:localCYgeo}).

Let us consider $S_1$. For this we change variables to $y=x-{1\over
2}(x_1+x_2)$ and the integral becomes
$$
S_1 ={g\over 2\pi }\int^{y_4}_{y_3}\sqrt{(y-y_3)(y-y_4)}\sqrt{y^2-z_1} \rd y.
$$
Expanding the second square root for $z_1$ small, each term in the series
can be computed explicitly and it is most easily given in terms of a generating
function \cite{civ},
\be\label{generone}
F(a) = -\pi \sqrt{(y_3+a)(y_4+a)}+{\pi\over 2}(y_3+y_4+2a)
\ee
as follows,
$$
S_1 = {g \over 32}(y_3+y_4)(y_4-y_3)^2 +{g\over 2\pi}\sum^{\infty}_{n=1}c_n
\Dd^{2n} F^{(n)}(0) 
$$
where $c_n$ are the coefficients in the expansion of $\sqrt{1-x}$ and
$F^{(n)}(a)$ is the $n$-th derivative with respect to $a$.

The explicit answer has the following structure,
\be
\label{funone}
S_1 = {g\over 4}z_2 I - {g\over {2I}}K(z_1,z_2,I^2),
\ee
where
$$
K(x,y,z) = \frac{1}{4}xy\left(1+{1\over 4z}(x+y)+{1\over
8z^2}(x+y)^2+{1\over 8z^2}xy+\ldots \right).
$$
It is important to notice that this is symmetric in $(x,y)$, namely,
$K(x,y,z)=K(y,x,z)$. This allows us to write,
\be\label{funtwo}
S_2 = -{g\over 4}z_1 I + {g\over {2I}}K(z_1,z_2,I^2).
\ee
In the following we will simplify the expressions by putting $ m=g=1$.
It will be useful to change variables to 
\be
\label{ts}
t=S_1 + S_2, \qquad s=\frac{1}{2}(S_1-S_2)
\ee
where $t$ is the total 't Hooft parameter.
Due to (\ref{funone}) and (\ref{funtwo}) one immediately obtains
\be\label{t2zexact}
t=\frac{1}{4}(z_2-z_1)\sqrt{1-2z_1-2z_2}.
\ee
Note, that $t$ can be regarded as a global parameter of the model. Different from $t$ the expression of $s$ in terms of the $z_i$ requires a transcendental function. This more complicated function reflects the dependence of $s$ on the choice of the symplectic basis in (\ref{symplectic}).

As mentioned earlier, there is another possibility to derive the periods as series in $z_i$ which was applied in \cite{hk}. There the authors consider a set of Picard--Fuchs differential operators, $\CL_1,\CL_2$ associated to the spectral curve and differential $\Omega$, which annihilate the periods. Therefore, 
these can be calculated as solutions to a system of ODEs. The Picard-Fuchs operators, which are given in eq.~(\ref{eq:PFop}) of appendix \ref{AppData}, have the following discriminant factors
\be
\text{disc}=z_1z_2 I^2 J=z_1z_2(1-2(z_1+z_2))(1-6z_1-6z_2+9z_1^2+14z_1z_2+9z_2^2).
\ee
Moreover, their solutions around $z_1=0$ and $z_2=0$ describe the periods of the elliptic geometry (\ref{ycurve}). Due to the fact that one can find a combination of periods such that the mirror map becomes exact (\ref{t2zexact}), it is convenient to introduce adapted coordinates $\tilde{z}_i$, $i=1,2$, by
\be
\tilde{z}_1=z_1+z_2,\quad \tilde{z}_2=\frac{1}{4}(z_1-z_2)\sqrt{1-2(z_1+z_2)},
\ee
as well as coordinates $\tilde{t}_{i}$, $i=1,2$, on the mirror by
\be
\tilde{t}_{1}=s=\frac{1}{2}(S_1-S_2),\quad \tilde{t}_{2}=t=S_1+S_2.
\ee

The Yukawa couplings may be found in eq.~(\ref{eq:YukFull}) as well as the genus one free energy $F_1$ in eq.~(\ref{eq:F1Full}). Due to the special type of the mirror map
\be
\tilde{z}_2=\tilde{t}_{2},
\ee
it is possible to derive a propagator which is of the following special form
\be
S=\begin{pmatrix} S^{\tilde{z}_1\tilde{z}_1} & 0 \\ 0 & 0 \end{pmatrix}.
\ee
For the technical details as well as for the ambiguities that have to be computed we refer the reader to appendix \ref{AppData}. With the help of this input it is easy to implement the direct integration procedure for the cubic matrix model as outlined in section \ref{sec:directint}. It turns out that we can recursively construct the free energies up to genus four. Moreover, we can also evaluate $F_g(S_1, S_2)$ for the cubic matrix model in perturbation theory, as was done in \cite{kmt,hk}. The expansions of our direct integration analysis read

\begin{footnotesize}
\be
\begin{split}
F_2&=-\frac{1}{240}\left(\frac{1}{S_1^2}+\frac{1}{S_2^2}\right)+\frac{35}{6}(S_1-S_2)+338S_1^2-1632S_1S_2+338S_2^2 %+\frac{66132}{5}S_1^3-120880S_1^2S_2+120880S_1S_2^2-\frac{66132}{5}S_2^3
+\CO(S^3)\\
F_3&=\frac{1}{1008}\left(\frac{1}{S_1^4}+\frac{1}{S_2^4}\right)+\frac{5005}{3}(S_1-S_2)+\frac{32}{9}\left(52522S_1^2-273403S_1S_2+52522S_2^2\right)
%+\frac{2}{7} \left(43194148 t[1]^3-427651275 t[1]^2 t[2]+427651275 t[1] t[2]^2-43194148 t[2]^3\right)
+\CO(S^3)\\
F_4&=-\frac{1}{1440}\left(\frac{1}{S_1^6}+\frac{1}{S_2^6}\right)+\frac{8083075}{6}(S_1-S_2)+\frac{880}{3}\left(788369S_1^2-4387436S_1S_2+788369S_2^2\right)
%+\frac{10}{9} \left(19854375338 t[1]^3-211076297241 t[1]^2 t[2]+211076297241 t[1] t[2]^2-19854375338 t[2]^3\right)
+\CO(S^3).
\end{split}
\ee
\end{footnotesize}
These results agree with the low-order results obtained in \cite{kmt,hk}. In the following we explain how to parameterize the ambiguity and how to fix the unknowns entering our ansatz.

\subsection{Direct integration, boundary conditions and integrability}
%----------------------------------
\label{sec:dirintboundaryintegrability}

In the last section we set up the necessary ingredients to perform a direct integration of the holomorphic anomaly equations. 
As mentioned in section \ref{sec:directint} the free energies $F_g$ can be written in the following way
\be
F_g=\sum_{k=1}^{3g-3}a_k(z_1,z_2)\left(S^{\tilde{z}_1\tilde{z}_1}\right)^k + f_g(z_1,z_2),
\ee
where $a_k$ are rational functions completely determined by the recursive procedure. $f_g$ is the holomorphic anomaly, which is not constrained by direct integration and must be fixed by supplying further boundary conditions. The amplitudes $F_g$ should be well-defined over the whole moduli space except for points at the boundary of moduli space where the elliptic geometry (\ref{ycurve}) acquires a node, i.e.~a cycle of ${\mathbb S}^1$-topology shrinks. Such points are known as conifold points and are given by the zero loci of the discriminant of the Picard--Fuchs system, which we also call conifold divisors.

Thus, regularity and holomorphicity imply that $f_g$ should be a rational function of $z_i$, where the numerator is at most of the same degree as the denominator. The denominator is given by the discriminant factors and takes the form $(z_1z_2J^2)^{2g-2}$. This gives the following ansatz for the holomorphic ambiguity
\be
f_g(z_1,z_2)=\frac{\sum_{k,l} a^{(g)}_{k,l} z_1^k z_2^l}{(z_1z_2 J^2)^{2g-2}},
\ee
where the $a^{(g)}_{k,l}$ have to be determined by the boundary conditions. Note that due to the symmetry of the model in $z_1$ and $z_2$ it is enough to restrict the numerator to a polynomial which is symmetric in $z_1$ and $z_2$. In order to be well defined as $z_i\rightarrow\infty$, the degree of this polynomial must be at most $12g-12$. It turns out that it is sufficient to truncate the degree at $9g-9$, as long as $g\leq4$. However, this reduced ansatz may not be present at higher genus and one would have to deal with the full ansatz of degree $12g-12$.

There are two boundary conditions which we will refer to in the following as $a$ cycle gap and $b$ cycle gap. Let us first consider the $a$ cycle gap. Due to the Gaussian contribution to the partition function of a multi-cut matrix model (see for example \cite{kmt} for more details) 
it is easily seen that the holomorphic expansion of $F_g$ at small filling fractions is of the form 
\be\label{Aconstraint}
F_g=\frac{B_{2g}}{2g(2g-2)}\left(\frac{1}{S_1^{2g-2}}+\frac{1}{S_2^{2g-2}}\right)+\CO(S),\qquad(g>1).
\ee
Due to the absence of subleading singular terms in $S_i$, this property of the expansion is referred to as the gap condition. The coefficients of the subleading singular powers of $S_i$ depend generically on the other $S_j$, with $j\neq i$ --in fact they are (infinite) series in the $S_j$. Demanding the vanishing of these series leads in principle to an over-determined system, therefore in the multi-parameter case it is not easy to count the number of independent conditions implied by (\ref{Aconstraint}).

The gap condition is also present in the expansion of genus $g$ topological string amplitudes near a conifold divisor \cite{alm,hkr,hkK3}, where we have
\be\label{Bconstraint}
F_g^c=\frac{B_{2g}}{2g(2g-2)\Pi^{2g-2}}+\CO(\Pi^0),\qquad(g>1).
\ee
Here, $\Pi$ is a flat coordinate normal to the divisor. In view of the Dijkgraaf--Vafa correspondence, this behavior should also characterize multi-cut matrix 
model amplitudes near the divisors of the spectral curve geometry. Again, since the coefficients of the subleading powers of $\Pi$ depend on the coordinates tangential to the conifold divisor, the counting of conditions in the multi-parameter case is not easily done.

However, it turns out that, when both constraints, (\ref{Aconstraint}) and (\ref{Bconstraint}), are taken into account, the holomorphic anomaly $f_g$ is completely and uniquely fixed. We checked this explicitly for genus $g\leq4$. It is then natural to conjecture that the $a$ and $b$ cycle gap conditions are always sufficient to fix all unknowns in the holomorphic ambiguity for general matrix models with polynomial potential. Following \cite{hkr} we refer to such a property as integrability of the holomorphic anomaly equation.

\subsection{Modular covariant formulation}
%-----------------------------------------
\label{sec:modularcovariant2}
In the last sections we explained how to solve the cubic matrix model with the techniques known from topological string theory. However, we used a somewhat artificial description which does not make the symmetry properties of the geometry completely explicit. Such a formulation is given by writing all quantities in a covariant modular way. Since the geometry is an elliptic curve together with meromorphic differential $\Omega$ we expect not only to parameterize the topological amplitudes $F_g$ by the elliptic modulus $\tau$ but in addition by an auxiliary parameter. In the following we explore how this can be achieved in detail.

We start by transforming the quartic curve (\ref{ycurve}) to Weierstrass form, where it is easy to read off the $j$-function. It is given by
\be
j(z_1,z_2)=\frac{16\left((1-3z_1-3z_2)^2+12z_1z_2\right)^3}{z_1z_2\left((1-3z_1-3z_2)^2-4z_1z_2\right)^2}.
\ee
Comparing this modular invariant to its usual Fourier expansion
\be
j(\tau)=q^{-1}+744+196884q+{\cal O}(q^2),
\ee
we get a relation $\tau=\tau(z_1,z_2)$. Using the definition of $j$ in terms of modular forms yields actually a rational expression.

It is also easy to identify the auxiliary parameter which accompanies $\tau$. Note that the periods/filling fractions are taken with respect to the differential $\Omega$, which is meromorphic. Thus the sum of all cycles is of course homologically trivial, but the sum of the periods does not have to vanish and is rather proportional to the residue of $\Omega$. Since this residue is related to the auxiliary parameter, it is natural to parameterize the topological amplitudes by both, $\tau$ and $t=g_s N$. Due to (\ref{t2zexact}) we obtain a relation $t=t(z_1,z_2)$.

In principle this allows us to rewrite all quantities in terms of modular forms together with an auxiliary parameter $t$, by combining the rational expression $\tau=\tau(z_1,z_2)$ with $t=t(z_1,z_2)$. If we do so, we obtain
\be
u=\frac{1-3z_1-3z_2}{2\sqrt{z_1z_2}},\qquad 4t=(z_1-z_2)\sqrt{1-2z_1-2z_2},
\ee
where $u$ is given in terms of modular forms $b$, $c$ and $d$ defined in appendix \ref{AppModEll} by
\be
u=\frac{c+d}{b}.
\ee
However, it turns out that, for the general cubic matrix model, the resulting formulae become too complicated. The reason is that the corresponding 
spectral curve is a generic elliptic curve. However, if we specialize the calculation to the slice $t=0$, or $S_1=-S_2$, the curve has $\Gamma(2)$ monodromy (it is the Seiberg--Witten curve of \cite{sw}) and it is possible to exploit the formulation in terms of modular forms, as we will see in section \ref{sec:slice}.

Fortunately, it is possible to give some closed and simple expressions using modular forms for the genus zero and one sectors, which will prove to be useful in due course. We start by quoting the perturbative calculation of the planar free energy \cite{civ}
\be
\ba
\label{planarfree}
F_0(S_1, S_2)&={1\over2}S_1^2\log\Bigl({S_1\over
m\Lambda^2}\Bigr) +{1\over2}S_2^2\log\Bigl({S_2\over
m\Lambda^2}\Bigr) -{3\over4}(S_1^2+S_2^2) +2S_1 S_2\log\Bigl({m\over
\Lambda g}\Bigr)\\ &\qquad +{1\over g\Delta^3}\Bigl({2\over3}S_1^3-5
S_1^2S_2+5 S_1 S_2^2 -{2\over3}S_2^3\Bigr)+{\cal O}(S^4).
\ea
\ee
Of course, $F_0(S_1, S_2)$ is symmetric under the exchange $S_1 \leftrightarrow -S_2.$ From the prepotential we can define the tau-coupling (\ref{tauij}) and also introduce
\be
2\pi \ri \tau={\partial^2 F_0 \over \partial s^2}.
\ee
It was shown in \cite{mswtwo} (see also \cite{bde}) that $\tau$ can be computed in terms of elliptic functions as 
\be
\label{eq:tauK}
\tau\, = \ri {\CK' \over \CK}\, = \ri {K(k')\over K(k)},
\ee
where
\begin{equation}\label{Kdef}
\ba
\CK &=  \int_{x_1}^{x_2} {\rd {z}\over\sqrt{|\sigma(z)|}} \,=\, {2\over \sqrt{{(x_1-x_3)(x_2-x_4)}}}\, K(k), \qquad  k^2 = {(x_1-x_2)(x_3-x_4)\over(x_1-x_3)(x_2-x_4)},\\
\CK' &=  \int_{x_2}^{x_3} {\rd {z}\over\sqrt{|\sigma(z)|}} = {2\over \sqrt{(x_1-x_3)(x_2-x_4)}}\, K(k'), \qquad k'^2 = 1-k^2.
\ea
\end{equation}
This modular parameter turns out to match with our definition by the $j$-function mentioned above. 
We find, in the full theory
\be\label{tautoS1S2}
\pi \ri \tau=\frac{1}{2}\log\Bigl( {-S_1 S_2 \over m^6}\Bigr) + {17 (S_1-S_2) \over m^3} + \frac{2 \left(83 S_1^2-209S_2 S-1+83 S_2^2\right)}{m^6}+\cdots
\ee
Let us now consider genus one. Akemann \cite{Akemann} gave a simple expression for $F_1$, that reads
\be\label{Fonematrix}
F_1=-{1\over24}\sum_{i=1}^4\ln M_i - \frac{1}{2}\ln K(k) -\frac{1}{12}\ln\Delta +{1\over8}\ln(a_1^{-}-a_2^{-})^2+{1\over8}\ln(a_1^{+}-a_2^{+})^2,
\ee
where $\Delta$ denotes the discriminant of $\sigma(x)$. Using that $M_i=g$ for the cubic matrix model as well as Thomae's formulae, cf.~app.~\ref{AppModEll}, this can be written compactly as 
\be
F_1=-\log\, \eta(\tau) - {1\over 24} \log \Delta,
\ee
where $\eta$ is the Dedekind eta-function.

%%%%%%%%%%%%%%%%%%%%%%%%%%%%%%%%%%%%%%%%%%%%%%%%%%%%%%%%%%%%%%%%%%%%%%%%%%%%
\sectiono{The cubic model for $S_1=-S_2$}\label{sec:slice}
In the following we will specialize the cubic matrix model studied in section \ref{sec:2cutcubicMM} to the slice $S_1=-S_2$. On this slice, $t=0$, and the direct integration procedure 
simplifies. Moreover, we are able to write all quantities which are needed for direct integration in terms of simple modular forms. The underlying reason for this is that, when $t=0$, 
the spectral curve of the matrix model becomes the Seiberg--Witten curve, which has simple monodromy properties. Therefore the recursive procedure will be very efficient in obtaining results at high genus. 

First of all notice that, by contour deformation, 
\be
S_1 + S_2=g \oint\limits_{z=0} {\rd z \over z^4} {\sqrt { 1+ {2m \over g} z + {m^2 \over g^2} z^2 + \frac{\lambda}{g^2} z^3 + \frac{\mu}{g^2} z^4}} 
={\lambda\over 2g}.
\ee
Therefore, if the parameter $\lambda$ in (\ref{splitf}) vanishes $\lambda=0$, we have
\be
t=S_1+S_2=0. 
\ee
In this case one also has \cite{dgkv}
\be
\tau_{11}=\tau_{22}=-\tau_{12}=\tau,
\ee
which can be seen from (\ref{tauij}).

From the point of view of the original matrix model, the slice $S_1=-S_2$ involves an analytic continuation in the space of 't Hooft parameters. This is because on this slice 
$S_1/S_2=N_1/N_2=-1$, which can not be implemented in the matrix integral (\ref{exmint}), since $N_{1,2}$ are {\it a priori} positive integers. In terms of matrix integrals, the slice $S_1=-S_2$ 
can be related to a {\it supermatrix model} \cite{agm,yost,dvdec,deynard}. A Hermitian supermatrix has the form
\be
\Phi=\begin{pmatrix} A & \Psi \\ \Psi^{\dagger} & C\end{pmatrix},
\ee
where $A$ ($C$) are $N_1\times N_1$ ($N_2 \times N_2$) Hermitian, Grassmann even matrices, and $\Psi$ is a matrix of complex, Grassmann odd numbers. 
The supermatrix model is defined by the partition function 
\be
Z_{\rm s} (N_1|N_2)=\int {\cal D} \Phi \, \re^{-{1\over g_s} {\rm Str} W(\Phi)},
\ee
where we consider a polynomial potential $W(\Phi)$ and ${\rm Str}$ denotes the supertrace. There are two types of supermatrix models with supergroup symmetry 
$U(N_1|N_2)$: the ordinary supermatrix model, and the physical supermatrix model \cite{yost}. The ordinary supermatrix model is obtained by 
requiring $A$, $C$ to be real Hermitian matrices, while the physical model is obtained by requiring that, after diagonalizing $\Phi$ by a superunitary 
transformation, the resulting eigenvalues are real. The partition function of the physical supermatrix model reads, in terms of eigenvalues \cite{yost,dvdec}
\be
\label{supermatrix}
Z_{\rm s} (N_1|N_2)={1\over N_1! N_2!} \int \prod_{i=1}^{N_1}\rd \mu_i \prod_{j=1}^{N_2} \rd \nu_j {\prod_{i<j} \left( \mu_i -\mu_j \right)^2 \left( \nu_i -\nu_j \right)^2 \over 
\prod_{i,j}  \left( \mu_i -\nu_j \right)^2} \re^{-{1\over g_s}\left(  \sum_i W(\mu_i) -\sum_j W(\nu_j)\right)},
\ee
where the two groups of eigenvalues $\mu_i$, $\nu_j$ are expanded around two different critical points of $W(x)$. This partition function is related to (\ref{exmint}) 
after changing $N_2 \rightarrow -N_2$ \cite{dvdec}, therefore it gives a physical realization of the $S_1/S_2<0$  slice of the moduli space. Notice, that the moduli space 
of the local Calabi--Yau for generic complex $S_1, S_2$ describes both the original matrix integral (\ref{exmint}) and its supergroup extension (\ref{supermatrix}). 

\subsection{The geometry}
In the following we discuss the geometry underlying the curve with $\lambda=0$. It is easy to see that, up to a shift in the $x$ coordinate, it can be written as 
\be
\label{scurve}
y^2=(x^2-a^2) (x^2-b^2), \qquad a>b. 
\ee
If we compare this to the Seiberg--Witten curve \cite{sw}
\be
y^2=(x^2-u)^2 -\Lambda^4_{\rm SW},
\ee
we find that they are equal once we identify the parameters as
\be
u={a^2 + b^2 \over 2} , \qquad \Lambda^2_{\rm SW}={a^2 -b^2\over 2}. 
\ee
We also want to translate these parameters in terms of the cubic matrix model variables. This was already done in \cite{dgkv,kmt}, and we have
\be
\Delta={m\over g}, \qquad u=\frac{1}{4} \Delta^2. 
\ee
We will set
\be
g=1.
\ee
On the other hand, we have the following relation between the $\Lambda$ parameter appearing in (\ref{planarfree}) and the Seiberg--Witten scale
\be
\Lambda={1\over {\sqrt{2}}} \Lambda_{\rm SW}. 
\ee

For the simple curve (\ref{scurve}) one can compute many quantities directly and relate them to modular forms or elliptic integrals. As a starting point the period integrals 
\be
S=S_1=-S_2, \qquad \Pi=\partial_{s} F_0(S,-S)
\ee
can be computed in terms of simple elliptic functions, which was done for $S$ in ref.~\cite{dgkv}. Repeating this analysis yields
\be
S ={1\over 2\pi \ri} \int_b^a  y(x) \rd x ={a\over 6\pi}\Bigl[ (a^2 + b^2) E(k_1) -2b^2 K(k_1)\Bigr]
\ee
as well as
\be\label{dualperiodslice}
\Pi= \int^b_{-b} y(x) \rd x =\frac{2}{3}a\Bigl[ (a^2 + b^2) E({k_1'}) +(b^2-a^2) K({k_1'})\Bigr],
\ee
where the elliptic modulus $k_1$ and its complementary one ${k_1'}$ are given by
\be
k_1^2={a^2-b^2\over a^2}, \qquad {k_1'}^2=1-k_1^2=\frac{b^2}{a^2}.
\ee
The modulus $k_1$ is related to the usual cross-ratio $k^2$ introduced in (\ref{Kdef}) as 
\be
k_1^2=\frac{4k}{(1+k)^2}.
\ee
In order to obtain expansions of the periods we introduce the parameters
\be
\mu={\Lambda_{\rm SW}^2 \over u}, \qquad \mu_{\rm D}=1-{\Lambda_{\rm SW}^2 \over u}.
\ee
Small $\mu$ corresponds to the semiclassical regime of Seiberg--Witten theory which occurs at $u\rightarrow\infty$, whereas small $\mu_{\rm D}$ relates to the region near $u\rightarrow\Lambda^2_{\rm SW}$, where a magnetic monopole becomes massless. In these variables the periods read
\be
\label{period}
{S\over u^{3\over 2}} = {{\sqrt{1+\mu}} \over 3\pi} \Bigl[ E\Bigl( {2\mu \over 1+\mu}\Bigr) +(\mu-1) K\Bigl( {2\mu \over 1+\mu}\Bigr)\Bigr],
\ee
\be
\label{dualperiod}
{\Pi\over u^{3\over 2}} = {4{\sqrt{2-\mu_{\rm D}}} \over 3} \Bigl[ E\Bigl( {\mu_{\rm D} \over 2-\mu_{\rm D}}\Bigr) +(\mu_{\rm D}-1) K\Bigl( {\mu_{\rm D} \over 2-\mu_{\rm D}}\Bigr)\Bigr].
\ee
Note that $S/u^{3/2}$ and $\Pi/u^{3/2}$ are dimensionless. Further, we expand (\ref{period}) around $\mu=0$ to obtain
\be
{S\over u^{3\over 2}} ={\mu^2 \over 8} + {3\mu^4 \over 256} + \frac{35 \mu^6}{8192}+\frac{1155\mu^8}{524288} + \frac{45045\mu^{10}}{33554432}+\cdots,
\ee
which is the expansion (4.19) of \cite{kmt}, after changing to the appropriate variables. The inverse expansion is given by
\be
\label{inversemu}
\mu^2=8 {S\over u^{3\over 2}} -6 \Bigl( {S\over u^{3/2}}\Bigr)^2 -{17\over 2} \Bigl( {S\over u^{3/2}}\Bigr)^3-\frac{375}{16}\Bigl( {S\over u^{3/2}}\Bigr)^4 - \frac{10689}{128}\Bigl( {S\over u^{3/2}}\Bigr)^5+\cdots.
\ee
We introduce now the following elliptic modulus $\tau_{0}$ as
\be
\label{taus}
\tau_{0}= \ri {K \Bigl( {1-\mu \over 1+\mu}\Bigr) \over K\Bigl( {2\mu \over 1+\mu}\Bigr)}=\ri \frac{K(k_1')}{K(k_1)}=\frac{\ri}{2}\frac{K(k')}{K(k)},
\ee
which can be expanded in $\mu$. By inverting this series one can derive $\mu$ as a function of $\tau_0$. In particular we observe
\be\label{muintau0}
\mu=\frac{b}{c+d},
\ee
where we follow the notation\footnote{~For our conventions on modular forms used in this section, see appendix \ref{AppModEll}.} of \cite{hk}.
In turn the expression (\ref{period}) defines the variable $\mu$ as a function of 
\be
\frac{S}{u^{3/2}}=8 \frac{S}{m^3}
\ee
 as well, 
and in particular the series (\ref{taus}) defines 
$\tau_0$ as a function of $S/u^{3/2}$:
\be
\label{tauSseries}
2\pi\ri\tau_{0} =\log \Bigl(\frac{S}{m^3}\Bigr) +34{S\over m^3} + 750 \Bigl( {S\over m^3}\Bigr)^2 + \frac{71260}{3} \Bigl({S\over m^3}\Bigr)^3 + \cdots.
\ee
Moreover, comparing with (\ref{tautoS1S2}) yields the identity
\be\label{tau0totau}
\tau_0=\frac{1}{2}\tau(S,-S),
\ee
which is obvious also from (\ref{taus}).

Consider now the dual elliptic modulus $\tau_{0,\rm D}$, obtained by a $S$-transformation on the elliptic modulus,
\be
\tau_{0,\rm D}=-\frac{1}{\tau_0}.
\ee
Following the same lines of thought as before, this defines $\tau_{0,\rm D}$ as a series in the dual period 
\be
\frac{S_{\rm D}}{u^{3/2}}=8 \frac{S_{\rm D}}{m^3}.
\ee

In the following we will set 
\be
\Lambda_{\rm SW}=1
\ee
so in particular $\mu=u^{-1}=4/m^2$. Note that (\ref{muintau0}) therefore defines $m$ as a function of $\tau_0$. Strictly speaking, $m$ is hence a function of $S$, but in order to establish the relation between $\tau_0$ and $\tau(S,-S)$, i.e.~(\ref{tau0totau}), we treated $m$ as an independent variable. In all subsequent formulas and expansions we will do so as well.

Next, we compute the Yukawa coupling
\be\label{Yukawacoupling}
C_{sss}={\partial^3 F_0 \over \partial s^3}.
\ee
This follows from the general formula for two-cut matrix models given by \cite{mswtwo} 
\be\label{fullthreef}
\frac{\partial^3 F_0}{\partial s^3} = \pi^3 \Bigl[M_1 \cdots M_4\, \CK^3\, \prod_{i<j} (x_i-x_j)^2\Bigr]^{-1}\cdot\sum_{i=1}^4\Bigl[\prod_{j\,\neq\, i}M_j\cdot \prod_{\stackrel{k,l\,\neq\, i,}{k<l}}(x_k-x_l)^2 \Bigr]
\ee
where $M_i=M(x_i)$, the spectral curve is written as in (\ref{scsigma}), and $\CK$ is given in (\ref{Kdef}) When applied to the Seiberg--Witten curve (\ref{scurve}) we obtain
\be\label{Yukawaex}
C_{sss}=\frac{\partial (4\pi\ri\tau_0)}{\partial s}=\frac{64\sqrt{2}}{m^3}\frac{(c+d)^{5/2}}{b^2cd}.
\ee
 To see this, one has to apply Thomae's formula, which relates the branch points $x_i$ to $\vartheta$-functions \cite{fay} and further one has to express $\CK$ in terms of modular forms. This is done as follows. Note that
\be
\CK=\int_{b}^{a}\frac{\rd x}{\sqrt{(a^2-x^2)(x^2-b^2)}}=\frac{1}{a}K(k_1).
\ee
The dimensionless combination $\sqrt{u}\,\CK$ can be expanded as a series in $\tau_0$ since
\be
\sqrt{u}\,\CK=\frac{1}{\sqrt{1+\mu}}K\left(\frac{2\mu}{1+\mu}\right).
\ee
This yields
\be
\CK=\frac{\pi}{m}\sqrt{\frac{c+d}{2}}.
\ee

We can check the formula (\ref{Yukawacoupling}) by calculating this quantity directly from the perturbative result. Evaluating the derivatives at $S_1=-S_2=S$ we obtain from (\ref{planarfree})
\be
\label{perturbativeyukawa}
{\partial^3 F_0 \over \partial s^3} ={2\over m^3} \Bigl\{ 34 +{m^3 \over S}  + 1500 {S \over m^3} + 71260 {S^2 \over m^6} +\cdots \Bigr\}.
\ee
Using (\ref{tauSseries}) this coincides with (\ref{Yukawacoupling}), if we treat $m$ as an independent variable.

The expression (\ref{Yukawaex}) for $C_{sss}$ is a modular form of weight $-3$ on the modular group $\Gamma(2)$ defined in the Appendix. We will use as generators for the 
ring of modular forms on $\Gamma(2)$, $M_{*}(\Gamma(2))$, the functions
\be
K_2=c+d,\qquad K_4=b^2,
\ee
which are modular forms of weight two and four, respectively. Note that instead of considering the $\Gamma(2)$ description of the Seiberg--Witten curve (\ref{scurve}) we could also use the equivalent $\Gamma_0(4)$ description, which amounts to trade $\tau_0$ for $2\tau_0=\tau(S,-S)$ in all expressions of this section.
\subsection{Direct integration and higher genus amplitudes}
Having discussed the genus zero sector of the cubic matrix model specialized to the slice $S_1=-S_2$, let us now turn our attention to the higher genus free energies $F_g$. According to \cite{hk,emo} the matrix model free energies $F_g$ can be promoted to modular invariant, non-holomorphic amplitudes $F_g(\tau_0,\bar\tau_0)$ which satisfy the holomorphic anomaly equations of \cite{bcov} in the local limit. The matrix model $F_g$ is recovered by formally considering the limit $\bar\tau_0\rightarrow\infty$ while keeping $\tau_0$ fixed. 

In order to apply this, we must compute the full non-holomorphic genus one amplitude $F_1$ and derive the propagator $S^{ss}$. Using the general formula (\ref{Fonematrix}) specialized to the Seiberg--Witten curve (\ref{scurve}), and by following the same argument as for the Yukawa coupling $C_{sss}$, we obtain
\be
F_1(\tau_0,\bar\tau_0)=-\log(\sqrt{\text{Im}\tau_0}\, \eta(\tau_0)\eta(-\bar\tau_0))+\frac{1}{4}\log\left(\frac{m^2 K_2}{\sqrt{K_4}}\right).
\ee
Indeed, when expanded we find
\be
F_1 = -\frac{1}{6}\log S +{S\over 3m^3} + 15 \left(\frac{S}{m^3}\right)^2 + {6202\over 9} \left(\frac{S}{m^3}\right)^3 + 32286 \left(\frac{S}{m^3}\right)^4 + \cdots,
\ee
which is precisely the series for $F_1$ obtained in \cite{kmt} after setting $S_1=-S_2=S$.

Next we turn to the propagator $S^{ss}$, defined by 
\be
\overline{C}^{ss}_{\bar s}=\bar{\partial}_{\bar s}S^{ss},
\ee
where $\overline{C}_{\bar s\bar s \bar s}$ is the complex conjugate of the Yukawa coupling $C_{sss}$ and the indices are raised by means of the metric
\be
G_{s\bar s}\sim \text{Im}\tau_0.
\ee
Using the chain rule and the relation (\ref{Yukawacoupling}) yields
\be
\partial_s F_1(\tau_0,\bar\tau_0)=-\frac{1}{48}C_{sss}\widehat{E}_2(\tau_0,\bar\tau_0)+\partial_s f_1(\tau_0),
\ee
where $f_1$ is given by
\be
f_1(\tau_0)=\frac{1}{4}\log\left(\frac{m^2 K_2}{\sqrt{K_4}}\right).
\ee
Hence, the propagator is identified with
\be
S^{ss}=-\frac{1}{24}\widehat{E}_2(\tau_0,\bar\tau_0).
\ee

Now we are prepared to apply the method of directly integrating the holomorphic anomaly equations according to \cite{gkmw, al}. In the conventions of this section the holomorphic anomaly equations can be cast into the following form
\be
\frac{\partial F_g}{\partial\widehat{E}_2}=-\frac{1}{192}C_{sss}^2\left[ \hat{D}_{\tau_0}^2 F_{g-1}+\frac{\hat{D}_{\tau_0}C_{sss}}{C_{sss}}\hat{D}_{\tau_0}F_{g-1}+\sum_{h=1}^{g-1}\hat{D}_{\tau_0}F_h \hat{D}_{\tau_0}F_{g-h}\right],\quad (g>1)
\ee
where $\hat{D}_{\tau_0}$ denotes the Maass derivative acting on (almost-holomorphic) modular forms of weight $k$ as
\be
\hat{D}_{\tau_0}=\frac{1}{2\pi\ri}\frac{\rd}{\rd\tau_0}-\frac{k}{4\pi\text{Im} \tau_0}.
\ee
Since the ring $\widehat{M}_{*}(\Gamma(2))=\IC[\widehat{E}_2,K_2,K_4]$ is closed under $\hat{D}_{\tau_0}$, and the $F_g$'s are modular invariant forms, the holomorphic anomaly equation can be integrated with respect to $\widehat{E}_2$. We obtain the following schematic result
\be\label{generalFg}
F_g(\tau_0,\bar\tau_0)=\widetilde\Delta^{2-2g}\cdot\sum_{k=1}^{3g-3}c_{k}^{(g)}(\tau_0)\widehat{E}_2^k(\tau_0,\bar\tau_0)+f_g(\tau_0),
\ee
where $c_{k}^{(g)}(\tau_0)$ are modular forms of weight $8(g-1)-2k$, completely determined by the holomorphic anomaly equation, and $\widetilde\Delta$ is just the denominator of $C_{sss}$. In particular it is a weight eight form given by
\be\label{DeltaTilde}
\widetilde\Delta=m^3(K_2^2-K_4)K_4.
\ee
All the non-trivial information is encoded in the holomorphic ambiguity $f_g(\tau_0)$. It has to be derived genus by genus by supplying further boundary conditions. In the particular case of the cubic matrix model specialized to the slice $S_1=-S_2$, we will argue in the next subsection that $f_g(\tau_0)$ can be fixed at all genera. Applying this procedure we were able to integrate the holomorphic anomaly equations and obtained the matrix model free energies to genus 52.

Let us at least present the result for the full non-holomorphic genus two amplitude
\be
\begin{split}
F_2(\tau_0&,\bar\tau_0)=
-\frac{160 K_2^5}{81m^6(K_2^2-K_4)^2 K_4^2}\widehat{E}_2^3
-\frac{16 K_2^4(5 K_2^2-7 K_4)}{9m^6(K_2^2-K_4)^2 K_4^2}\widehat{E}_2^2\\
&-\frac{8 K_2^3 (77 K_2^4-132 K_2^2 K_4+63 K_4^2)}{27m^6(K_2^2-K_4)^2 K_4^2}\widehat{E}_2
-\frac{4K_2^4(2051K_2^4-4005K_2^2K_4+1890K_4^2)}{405m^6(K_2^2-K_4)^2 K_4^2}.
\end{split}
\ee

Here we collect some low genus expansions of the free energy amplitudes of the cubic matrix model on the slice $S_1=-S_2=S$:

\begin{footnotesize}
\be
\begin{split}
m^6 F_2& =-\frac{1}{120}\frac{m^6}{S^2}+\frac{35}{3}\frac{S}{m^3}+2308\frac{S^2}{m^6}+\frac{1341064}{5}\frac{S^3}{m^9}+24734074\frac{S^4}{m^{12}}+\cdots\\
m^{12} F_3& =\frac{1}{504}\frac{m^{12}}{S^4}+\frac{10010}{3}\frac{S}{m^3}+\frac{4036768}{3}\frac{S^2}{m^6}+\frac{1883381692}{7}\frac{S^3}{m^9}+38608040638\frac{S^4}{m^{12}}+\cdots\\
m^{18} F_4& =-\frac{1}{720}\frac{m^{18}}{S^6}+\frac{8083075}{3}\frac{S}{m^3}+1749491040\frac{S^2}{m^6}+\frac{4618613451580}{9}\frac{S^3}{m^9}+\cdots\\
m^{24} F_5& =\frac{1}{528}\frac{m^{24}}{S^8}+\frac{13013750750}{3}\frac{S}{m^3}+4038280413440\frac{S^2}{m^6}+\frac{17515677810823140}{11}\frac{S^3}{m^9}+\cdots\\
m^{30} F_6& =-\frac{691}{163800}\frac{m^{30}}{S^{10}}+11699361924250\frac{S}{m^3}+\frac{43710230883020800}{3}\frac{S^2}{m^6}+\cdots.
\end{split}
\ee
\end{footnotesize}

We can check some of these results by comparing to the perturbative calculations of \cite{kmt} specialized to $S_1=-S_2=S$. We observe agreement for genus two and three at low order in $S/m^3$. All higher genus computations are new results.

The direct integration procedure outlined here is by far the most efficient method to calculate higher genus amplitudes in matrix models. It only takes a few minutes to reach e.g.~genus 10 on a conventional personal computer.

\subsection{Boundary conditions and integrability}
%--------------------------------------------------
\label{sec:boundarycondandint} 
According to \cite{abk, emo} $F_g$ is an almost-holomorphic modular invariant form under the spacetime duality group, in this case $\Gamma(2)$. Hence, $F_g$ is regular except for some points on the boundary of moduli space.

Regularity and holomorphicity imply that $f_g$ should be a rational function, where its denominator is given by an appropriate power of the discriminant of the curve. From the expression (\ref{generalFg}) we see that the denominator of $f_g$ is given by $\widetilde\Delta^{2g-2}$, hence a weight $8(g-1)$ form. Modularity now implies that the numerator has to be a form of finite weight, in order to cancel the weight from the denominator. Since the space of weight $k$ forms is finite dimensional, there are only finitely many coefficients to determine. In particular, for $\Gamma(2)$ we have
\be
\text{dim}\, M_k(\Gamma(2))=\begin{cases}\frac{k+2}{2},& k>2,\quad k\text{ even.}\\0,&\text{else.}\end{cases}
\ee
In summary this justifies the ansatz
\be
f_g(\tau_0)=\widetilde\Delta^{2-2g}\cdot\sum\limits_{k=0}^{4(g-1)}a_kK_2^{2k}K_4^{4(g-1)-k},\qquad (g>1)
\ee
where $\widetilde\Delta$ is given in eq.~(\ref{DeltaTilde}). This implies that there are $4g-3$ unknown constants $a_k$ in the ambiguity $f_g$. These are completely and uniquely fixed by imposing the following two boundary conditions.

First, we know that the holomorphic expansion of $F_g$ at small $S$ has the structure (\ref{Aconstraint}) specialized to the slice, which imposes $2g-1$ conditions on $f_g$ and leaving $2g-2$ unknowns. Further the holomorphic expansion at conifold divisors is of the form (\ref{Bconstraint}), where $\Pi$ is a suitable coordinate transverse to the divisor which vanishes at the conifold. In our case $\Pi$ is the dual period. Thus, (\ref{Bconstraint}) imposes $2g-2$ further constraints on the ambiguity, and it determines it completely.

%%%%%%%%%%%%%%%%%%%%%%%%%%%%%%%%%%%%%%%%%%%%%%%%%%%%%%%%%%%%%%%%%%%%%%%%%%%%
\sectiono{Non-perturbative aspects}
\label{sec:nonpert2cutmm}

In this section we address non-perturbative effects in the two-cut matrix model, and its connection to the large order behavior of the $1/N$ expansion. 
We first review the one-cut case. 

\subsection{Non-perturbative effects in the one-cut matrix model}
%-----------------------------------------------------------------
\label{sec:onecutmatrixmodel}  

For concreteness, we will focus here on the cubic matrix model which we are analyzing in this paper. 

In the one-cut cubic matrix model, the large $N$ limit is described by a distribution of eigenvalues around the 
minimum of the potential at $x=0$. The eigenvalues fill the interval $[a,b]$. It has been known for some time that instanton sectors in this model are obtained by tunneling a 
finite, small number of eigenvalues $\ell \ll N$ from this interval to the maximum of the effective potential, located at $x_0$. The structure of the 
partition function in the $\ell$-instanton sector has been determined in \cite{mswone,mswtwo}, and at one loop it has the form
\be
Z^{(\ell)} = {g_s^{\ell^2/2} \over (2\pi)^{\ell/2}}\, G_2(\ell+1) \mu_1^{\ell^2}\, \exp \left( - \frac{\ell A}{g_s} \right) \biggl\{ 1 +  \CO(g_s)\biggr\}.
\ee
In this equation, $G_2(z)$ is the Barnes function. $A$ is the instanton action, and it can be computed in terms of the spectral curve of the one-cut matrix model as
\be
A=\int_b^{x_0} \rd z \, y(z).
\ee
Finally, $\mu_1$ is the one-loop contribution, and it has the explicit expression 
\be
\mu_1={b-a\over 4} {1\over {\sqrt{M(x_0) [(a-x_0)(b-x_0)]^{5\over 2}}}}.
\ee
In \cite{mswone} it was argued, following standard arguments in the large order behavior of perturbation theory \cite{zj}, 
that the free energy of the one-instanton amplitude, $F^{(1)}$, should determine the leading asymptotics at large $g$ of the perturbative amplitudes $F_g$, according  
to the formula 
\be
\label{dispersion}
F_g = {1\over 2\pi \ri} \int_0^{\infty} {\rd z  \over z^{g+1}}F^{(1)}(z).
\ee
If we write
\be
\label{oneinstanton}
F^{(1)}= g_s^{1/ 2} \re^{-A/g_s} \sum_{\ell=1}^{\infty} \mu_l g_s^{\ell-1}, 
\ee
we obtain the full $1/g$ asymptotics
\be
\label{onecutas}
F_g \sim_g {1\over \pi} A^{-2g-b} \Gamma\Bigl(2g+b \Bigr)\sum_{\ell=1}^{\infty} { \mu_\ell A^{\ell-1} \over\prod_{k=1}^{\ell-1} (2g+b-k) }.
\ee
where
\be
\label{bgrav}
b=-{5\over 2}.
\ee
The formula (\ref{onecutas}) can be regarded as a generalization of the 
asymptotics for formal solutions of nonlinear ODEs. The reason is as follows. In the double-scaling limit of the matrix model 
(see \cite{dfgzj}), the total free energy of the matrix model becomes 
a function of a double-scaled variable $z$, 
\be
F(t,g_s) \rightarrow F_{\rm ds}(z), 
\ee
and the specific heat $u=-F''_{\rm ds}(z)$ satisfies the Painlev\'e I equation 
\be
u^2-{1\over 6} u''=z.
\ee
In particular, the genus expansion of the cubic matrix model leads to a formal solution of Painlev\'e I 
\be
\label{panex}
u(z)=z^{1/2} \sum_{g=0}^{\infty} u_{g,0} z^{-5g/2}. 
\ee
On the other hand, the instanton sectors of the matrix model lead to instanton corrections of the form 
\be
\label{pinstantons}
u_\ell(z)=z^{1/2-5\ell/8} \re^{-\ell a  z^{5/4}}\sum_{n=0}^{\infty} u_{n,\ell} z^{-5n/4}
\ee
where
\be
a={8 {\sqrt{3}} \over 5}.
\ee
It can be shown that the coefficients of (\ref{panex}) have an asymptotic behavior at large $g$ which is governed by the one-instanton solution $u_1(z)$ in (\ref{pinstantons}). 
The precise formula is,
\be
\label{ug0}
u_{g,0} \sim_g  {a^{-2g+{1\over 2}} \over \pi}\, \Gamma\Bigl(2g-{1\over 2} \Bigr)\, 
{{\cal S}_1 \over \pi \ri} \biggl\{1 + \sum_{l=1}^{\infty} {u_{l,1} a^{l} 
\over \prod_{k=1}^{l} (g-1/2 -k)} \biggr\},
\ee
where ${\cal S}_1$ is a Stokes constant. One can explicitly check \cite{david,mswone} that (\ref{ug0}) can be deduced from the double-scaling limit of 
the asymptotics (\ref{onecutas}). In particular, the constant $a$ is the double-scaling limit of the instanton action.

\subsection{Non-perturbative effects in the cubic matrix model}
%---------------------------------------
\label{sec:nonperturbativecubicmm}

Non-perturbative effects in multi-cut matrix models have been studied in \cite{bde, mswtwo}. A multi-cut matrix model with a fixed choice of filling fractions 
must be regarded as a fixed background, and any other choice of filling fractions leads to an instanton correction to the free energy on the fixed 
background. To be concrete, let us 
consider a two-cut matrix model with a fixed background given by $N_1$, $N_2$ eigenvalues in the stable and unstable saddle points, respectively. 
The partial 't Hooft parameters $S_1$, $S_2$ are given as usual by 
$S_i=g_s N_i$. The total partition function is of the form 
\be
Z=Z(N_1, N_2) + \sum_{\ell\not=0} \zeta^{\ell} \, Z(N_1-\ell, N_2+\ell). 
\ee
The sum over $\ell$ corresponds to the tunneling of $\ell$ eigenvalues from the first cut to the second cut, and 
at large $N$, the corresponding partition functions have the form 
\be
\label{zell}
Z^{(\ell)} = \zeta^{\ell} q^{\ell^2/2}  \exp \left( - \frac{\ell A}{g_s} \right) \biggl\{ 1 +  \CO(g_s)\biggr\}, \qquad \ell \in \IZ^*
\ee
where 
\be
A=\partial_{s} F_0 \qquad \mathrm{and} \qquad q=\exp\Bigl( \partial_s^2 F_0 \Bigr).
\ee
The variable $s$ is given in (\ref{ts}). If the cuts of the matrix model are the intervals $[x_1, x_2]$, $[x_3, x_4]$, the instanton action $A$ can be written as 
\be
\label{2cutaction}
A=\int_{x_2}^{x_3}  y(x) \rd x. 
\ee
If ${\rm Re}(A) \not= 0$, the instanton contributions are exponentially 
suppressed if ${\rm sgn}({\rm Re}(A)\ell )>0$, and they are exponentially enhanced if ${\rm sgn}({\rm Re}(A)\ell )<0$. This is just reflecting the fact that 
the generic background is unstable and if we expand around it we will find tachyonic directions. Notice however that both 
corrections are non-perturbative in $g_s$, therefore they are invisible in the genus expansion. 

It is generically expected that the existence of these non-perturbative sectors leads to the factorial divergence of the genus expansion around a fixed background.  The growth of the perturbative string amplitudes at large genus (and fixed $S_1$, $S_2$) should be of the same form as in (\ref{onecutas}), i.e. 
\be
\label{roughas}
F_g(S_1, S_2) \sim_g A^{-2g-b}\, \Gamma(2g+b) + \CO(g^{-1})
\ee
where $A$ is given by (\ref{2cutaction}) and $b$ is a constant.

We can test these predictions by numerical methods using our results from direct integration. We start by concentrating on the slice $S_1=-S_2$. Note that in this case the instanton action $A$ is given by the dual period $\Pi$, whose explicit expression is given in eq.~(\ref{dualperiodslice}). In order to extract the asymptotic of the sequence $\{F_g\}_{g\geq0}$ we employ a standard numerical technique known as Richardson extrapolation. The method removes the first terms of the subleading tail and hence accelerates the convergence. Given a sequence $\{S_g\}_{g\geq0}$ in the form
\be\label{richard}
S_g=a_0+\frac{a_1}{g}+\frac{a_2}{g^2}+\dots
\ee
its Richardson transform is defined by
\be
R_S(g,N)=\sum_{k\geq0}\frac{(-1)^{k+N}(g+k)^N}{k!\, (N-k)!}S_{g+k},
\ee
such that the sub-leading terms in $\{S_g\}_{g\geq0}$ are cancelled up to order $g^{-N}$. In fact, it can be shown that if $\{S_g\}_{g\geq0}$ is a finite sequence, the Richardson transform returns exactly the leading term $a_0$.

\begin{figure}[!ht]
\leavevmode
\begin{minipage}{0.5\textwidth}
\begin{center}
\epsfysize=5cm
\epsfbox{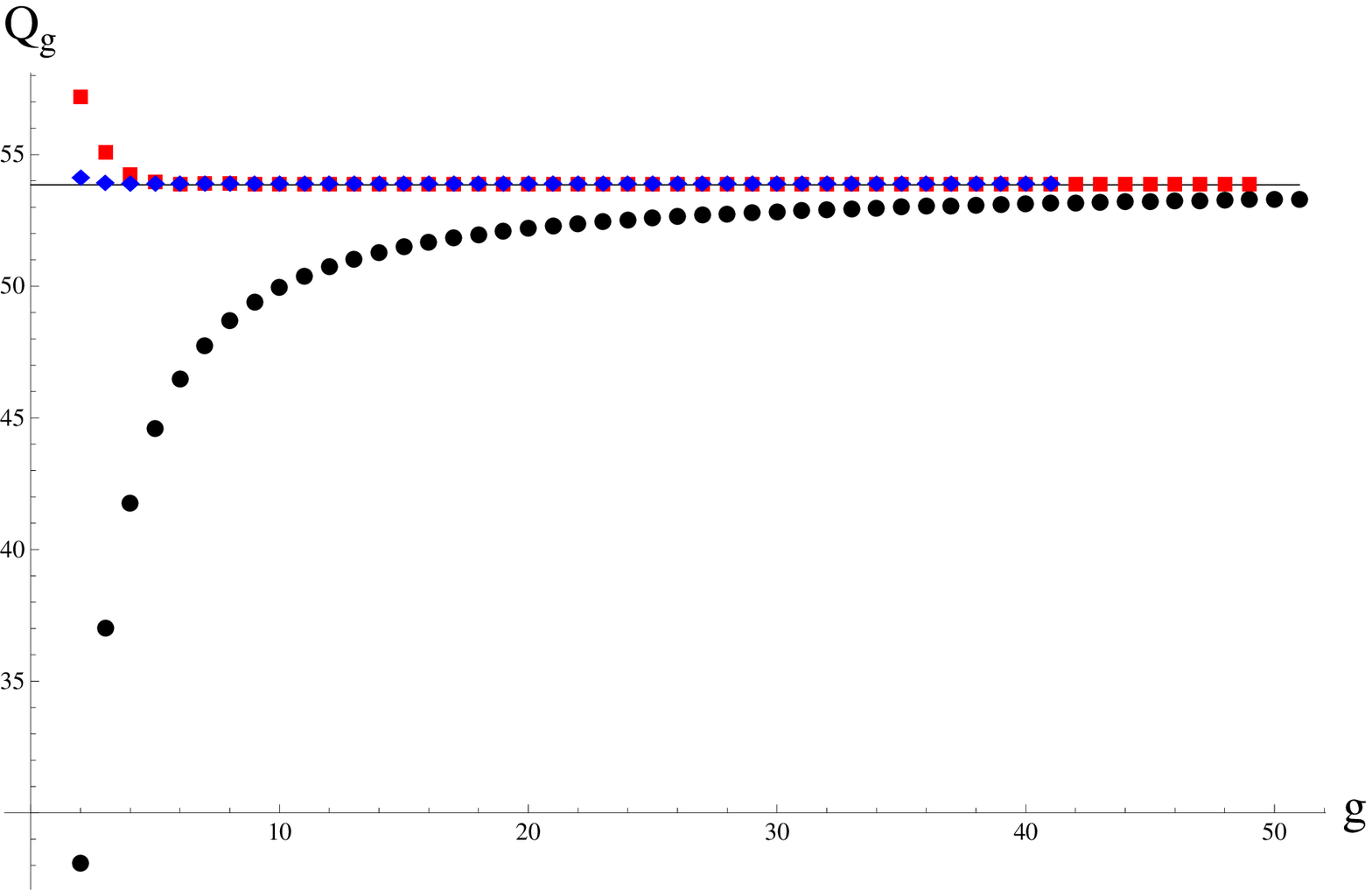}
\end{center}
\end{minipage}
\begin{minipage}{0.5\textwidth}
\begin{center}
\epsfysize=5cm
\epsfbox{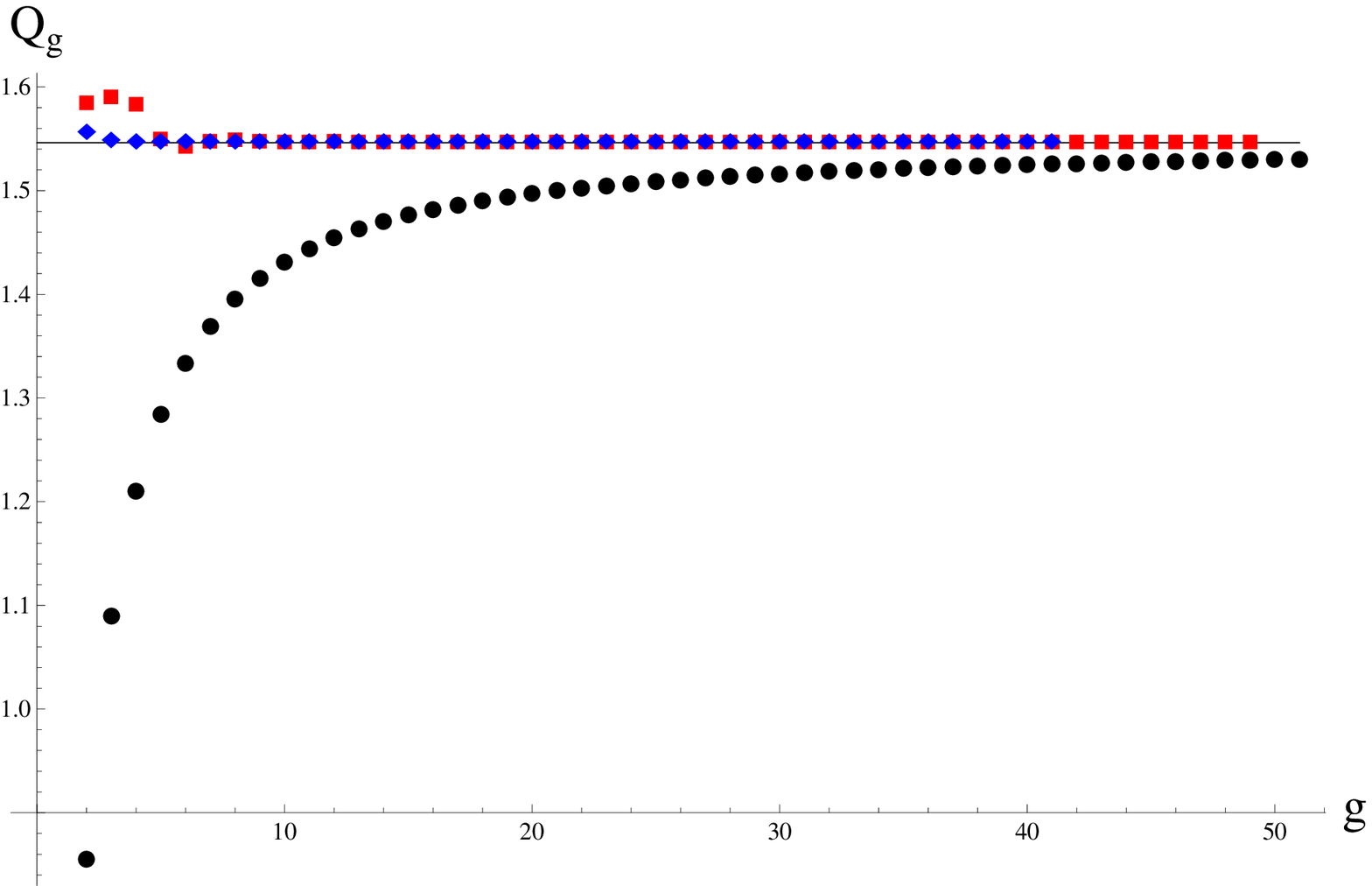}
\end{center}
\end{minipage}
\caption{The sequence $Q_g$ (\textbullet) and two Richardson transforms (\textcolor{red}{$\blacksquare$}, \textcolor{blue}{$\blacklozenge$}) at $\tau_0=\frac{\ri}{2}$ (left) and $\tau_0=\frac{2\ri}{3}+\frac{1}{9}$ (right) which corresponds to $S\approx 0.139$ and $S\approx 0.117+0.016\ri$, respectively. The leading asymptotics as predicted by the instanton action $|A|^{-2}$ is shown as a straight line. The error for genus 52 is about $10^{-8}$ \% and $10^{-10}$ \%, resp.}
\label{LOQfig}
\end{figure}

\begin{figure}[!ht]
\leavevmode
\begin{minipage}{0.5\textwidth}
\begin{center}
\epsfysize=5cm
\epsfbox{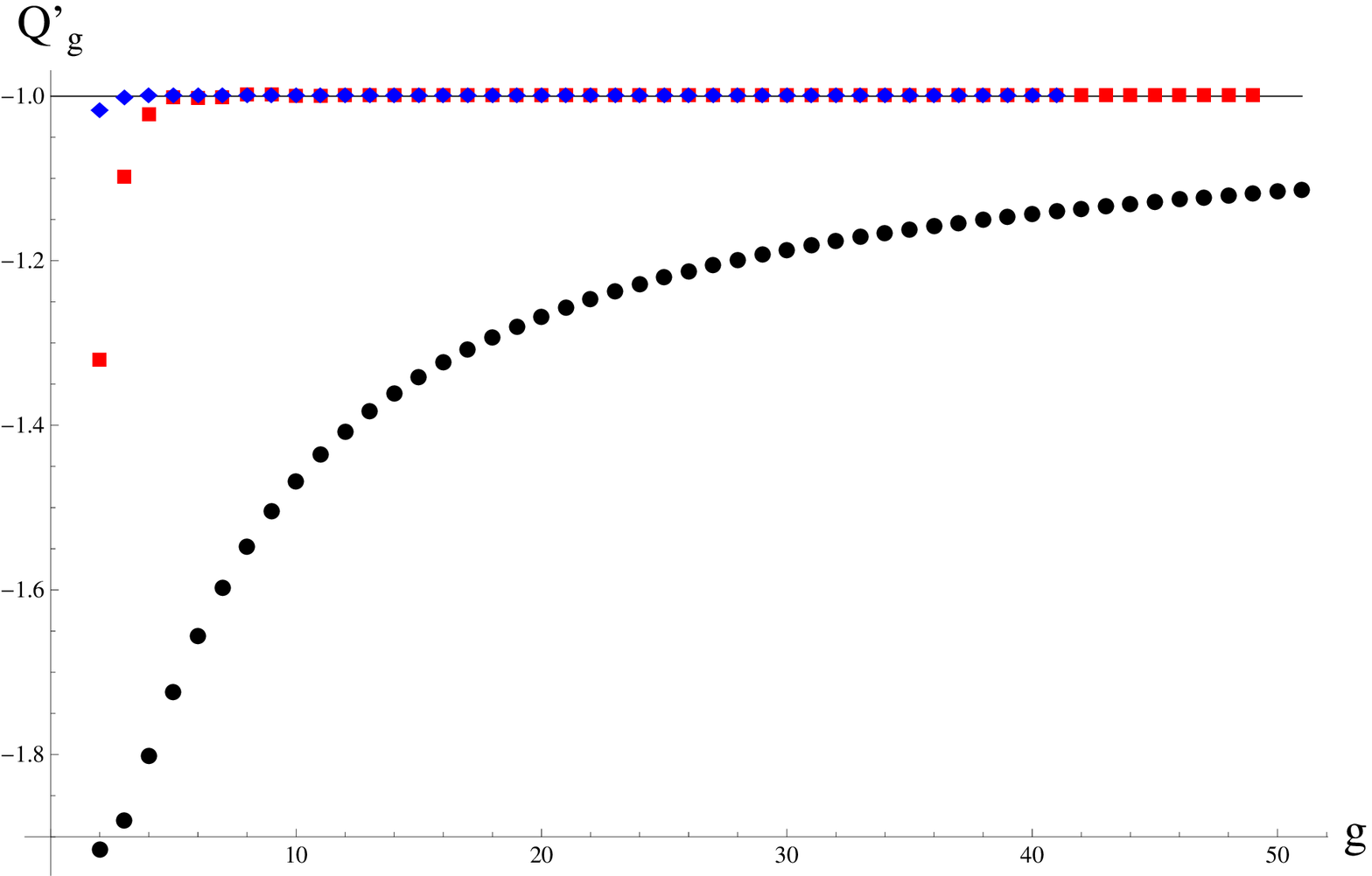}
\end{center}
\end{minipage}
\begin{minipage}{0.5\textwidth}
\begin{center}
\epsfysize=5cm
\epsfbox{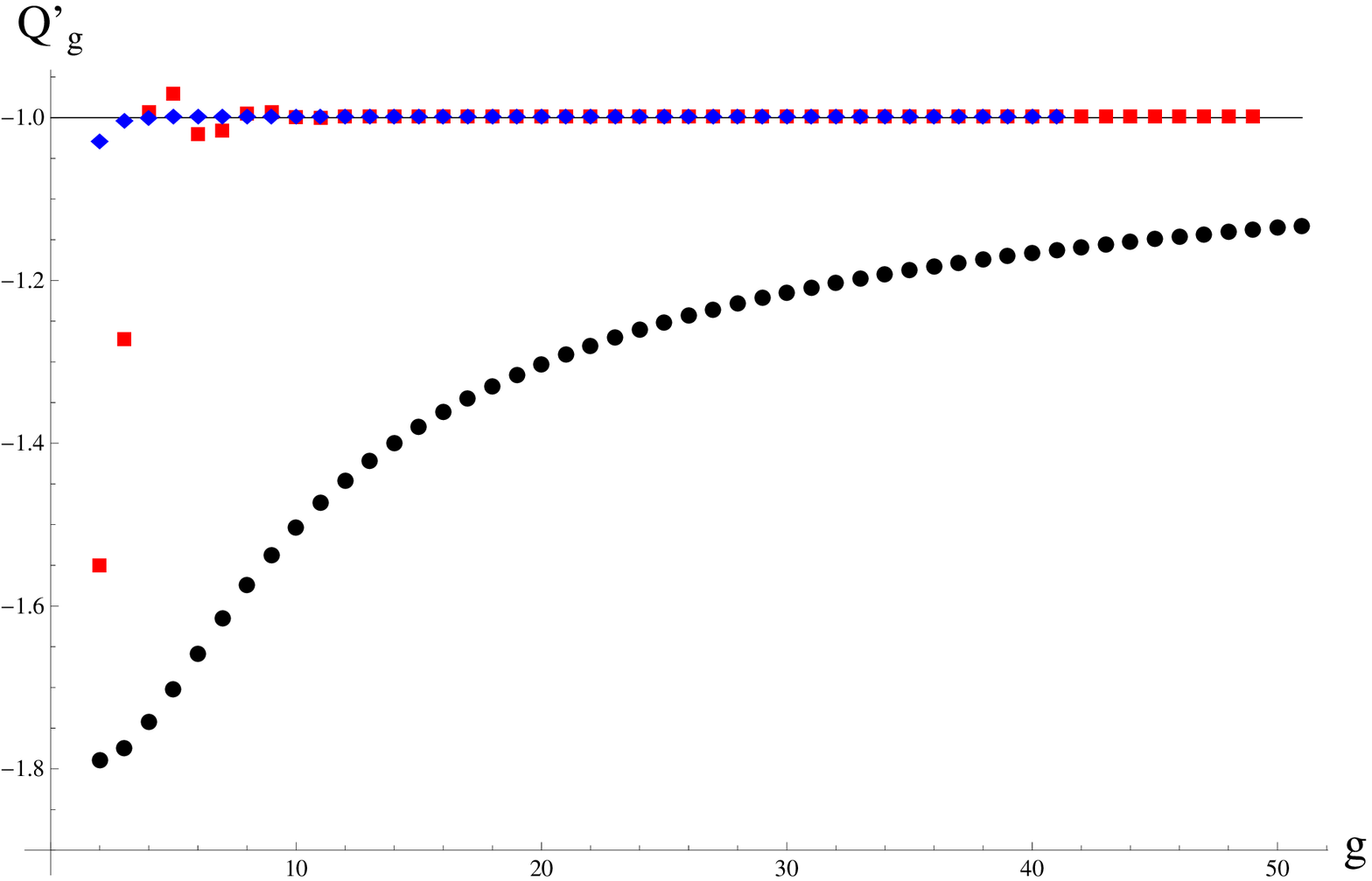}
\end{center}
\end{minipage}
\caption{The sequence $Q'_g$ (\textbullet) and two Richardson transforms (\textcolor{red}{$\blacksquare$}, \textcolor{blue}{$\blacklozenge$}) at $\tau_0=\frac{\ri}{2}$ (left) and $\tau_0=\frac{2\ri}{3}+\frac{1}{9}$ (right) which corresponds to $S\approx 0.139$ and $S\approx 0.117+0.016\ri$, respectively. The leading asymptotics as predicted by the parameter $b=-1$ is shown as a straight line. The error for genus 52 is about $10^{-8}$ \% in both cases.}
\label{LOQfig2}
\end{figure}

Comparing (\ref{richard}) with (\ref{roughas}) one can extract the instanton action by considering the sequence
\be\label{Qgseries}
Q_g=\frac{F_{g+1}}{4g^2 F_g}=\frac{1}{A^2}\left(1+\frac{1+2b}{2g}+\CO(g^{-2})\right).
\ee
Once $A$ is confirmed, one can then obtain the parameter $b$ from the new sequence
\be
Q'_g=2g\left(A^2 Q_g-1\right)=1+2b+\CO(g^{-1}).
\ee
In Fig.~\ref{LOQfig} and Fig.~\ref{LOQfig2} we plot the sequences $Q_g$, $Q'_g$, together with their Richardson transforms, for 
two values of $S$. It is obvious from the numerical calculation that the large genus asymptotics is controlled at leading order by the instanton action. 
In addition, we find numerically that
\be
\label{bone}
b=-1. 
\ee
This value of $b$ is different from the one characterizing the one-cut model (\ref{bgrav}). In fact, the value (\ref{bgrav}) corresponds to the universality class 
of pure two-dimensional gravity, while the value (\ref{bone}) corresponds rather to the universality class of the $c=1$ string \cite{ps}. It is interesting to see that 
both behaviors are present in the two-cut cubic matrix model, along different submanifolds of the moduli space (the 2d gravity behavior takes place in the slice $S_2=0$, while the $c=1$ behavior takes place in the slice $S_1+S_2=0$).  

\begin{figure}[!ht]
\leavevmode
\begin{center}
\epsfysize=7cm
\epsfbox{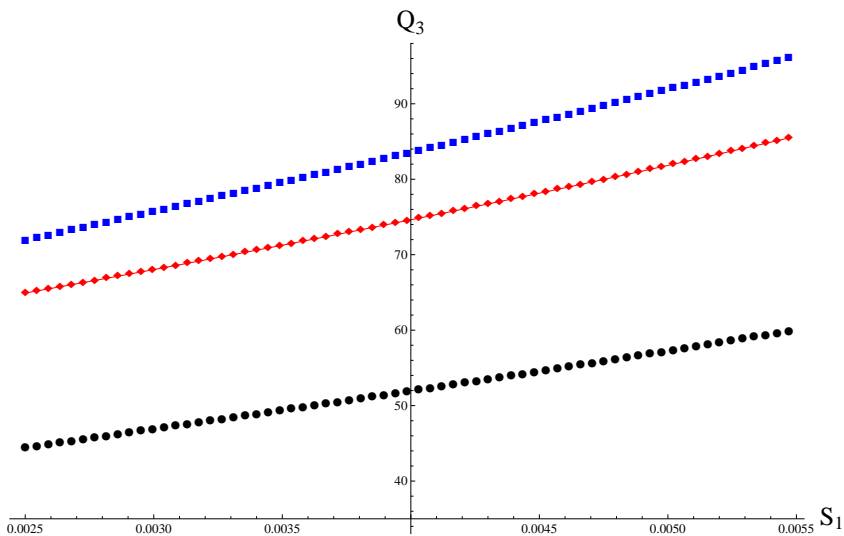}
\end{center}
\caption{$Q_3$ (\textbullet), $R_Q(1,2)$ (\textcolor{blue}{$\blacksquare$}) and $|A|^{-2}$ (\textcolor{red}{$\blacklozenge$}) are plotted for several values of $S_1$ around the slice point $S_1=-S_2=S=0.004$. $Q_3$ and $R_Q(1,2)$ have a relative error of about $30$ \% and $10$ \%, respectively, as compared to the instanton action $|A|^{-2}$ throughout the data set.}
\label{LOQfig3}
\end{figure}

Turning our attention to a generic value of the filling fractions $(S_1,S_2)$ in the cubic matrix model, we can try to test our prediction (\ref{roughas}) by using the results from direct integration of section \ref{sec:2cutcubicMM}. Since we computed $F_g$ up to genus four, we can only explore the first four elements of the sequence $\{Q_g\}_{g\geq0}$, eq.~(\ref{Qgseries}). In order to have a better control of the error, we consider a perturbation around the submanifold $S_1+S_2=0$ where the large order behavior is well established. The instanton action (\ref{2cutaction}) is calculated using (\ref{periodsapp}) by
\be
\begin{split}
A&=\frac{\p F_0}{\p S_1}-\frac{\p F_0}{\p S_2}=\Pi_1-\Pi_2\\
&=\log(S_1)S_1-\log(S_2)S_2+\frac{1}{6}-S_1+S_2+\CO(S^2).
\end{split}
\ee
Fig.~\ref{LOQfig3} shows $Q_3$, $R_Q(1,2)$ and $|A|^{-2}$ as a function of $S_1$ in the vicinity of the slice point $S_1=-S_2=S=0.004$, where convergence is ensured. We observe that the behavior of $Q_3$ and $R_Q(1,2)$ is qualitatively the same as predicted by the instanton action. Moreover, their relative errors stay roughly constant over the complete data set. This seems to indicate that the large order behavior of the genus expansion is also governed by the instanton action in the general two-cut cubic matrix model.

Unfortunately, our numerical results for the generic case are not good enough to determine the value of $b$ reliably. It is an interesting question to know 
how this value changes as we move in the moduli space. We expect it to be $b=-1$ except in the one-cut slices $S_1=0$, $S_2=0$, where it takes the value 
(\ref{bgrav}). 

\subsection{Asymptotics and non-perturbative sectors}
\label{sec:asymptotics}

 In principle, one should be able to refine the asymptotic formula (\ref{roughas}) and obtain a generalization of 
(\ref{onecutas}) involving the $g_s$ expansion of instanton solutions. A natural guess is that the relevant instanton solutions are the closest ones to the given background, 
i.e. the instanton amplitudes (\ref{zell}) with $\ell=\pm1$. This guess would relate the large genus behavior of $F_g(t_1, t_2)$ to an integral of the form (\ref{dispersion}), involving this time $F^{(1)}$ and $F^{(-1)}$. However, this expectation turns out to be too naive. Indeed, it seems that the asymptotics involves 
{\it new non-perturbative sectors} whose matrix model interpretation is yet unknown. 

\begin{figure}[!ht]
\leavevmode
\begin{center}
\epsfysize=7cm
\epsfbox{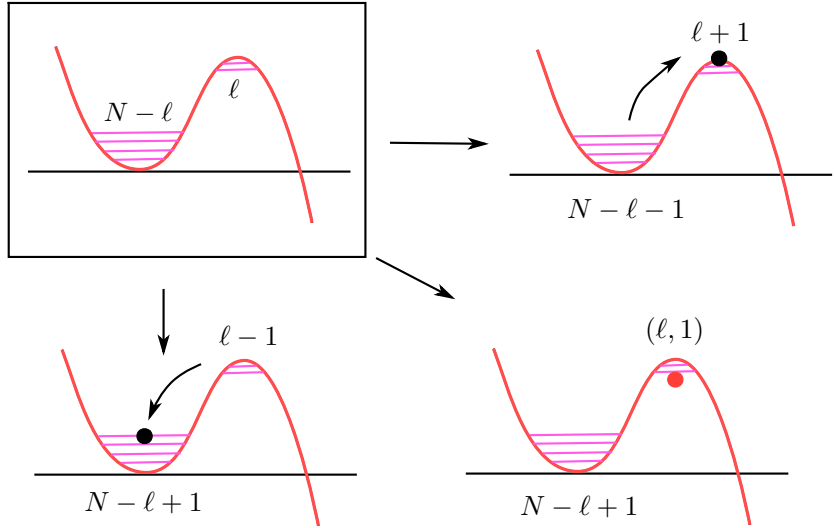}
\end{center}
\caption{The asymptotics of the coefficients of the $\ell$-th instanton solution $u_{\ell}(z)$ of Painlev\'e I 
is determined by the two nearest neighbor instantons, which are obtained by eigenvalue tunneling, and by 
the generalized instanton amplitude $u_{\ell|1}$, which is represented here by the label $(\ell,1)$.}
\label{painlevefig}
\end{figure} 

In order to explain this in some detail, we will come back to a simplest case where the asymptotics 
can be fully determined, namely the Painlev\'e I equation and its instanton solutions 
$u_{\ell}(z)$. It is natural to ask what is the asymptotics of the coefficients $u_{n,\ell}$ appearing in (\ref{pinstantons}). Notice that, 
when $\ell$ is big, this instanton solution is the 
double-scaled limit of a two-cut solution, therefore the question of the asymptotics of this sequence is closely related to the original question concerning the asymptotics (\ref{roughas}). 
We have seen in (\ref{ug0}) that the asymptotics of the perturbative solution is governed by the one-instanton solution. In the same way, one would think that the asymptotics of 
the $\ell$-instanton solution is governed by the $\ell\pm 1$ instanton amplitudes. It has been shown in \cite{gikm} that this is not the case. 
In order to understand the asymptotics of a generic instanton sector, one has to consider more general 
amplitudes, labelled by two non-negative integers:
\be
u_{n|m}(z).
\ee
The amplitude where $m=0$ is the standard instanton amplitude: $u_{\ell|0}(z)=u_{\ell}(z)$. The other amplitudes can be obtained by requiring 
\be
\label{ccts}
u(z, C_1, C_2)=\sum_{n,m\ge 0} u_{n|m}(z) C_1^n C_2^m 
\ee
to be a formal solution to the Painlev\'e I equation, for arbitrary $C_1, C_2$, and that 
\be
u_{n|m}(z) \sim \re^{-(n-m)a z^{5/4}}, \qquad z\rightarrow \infty. 
\ee
The two-parameter solution of the Painlev\'e I equation (\ref{ccts}) is called a {\it trans-series solution}, and it was introduced by Jean \'Ecalle in the 
context of resurgent analysis (see for example \cite{ss} for an simple introduction to resurgence). 
It turns out that the asymptotic behavior of the coefficients $u_{n,\ell}$ in the $\ell$-th instanton $u_{\ell}$ is governed by the solutions $u_{\ell\pm1}(z)$, 
but also by the solution $u_{\ell|1}(z)$. This means that the asymptotics of the coefficients $u_{n,\ell}$ as $n\rightarrow \infty$ can be obtained by a relation similar to (\ref{dispersion}), but involving $u_{\ell\pm1}(z)$ as well as $u_{\ell|1}(z)$. For example, let us consider the one-instanton solution, and let us ask what is 
the asymptotics of the coefficients $u_{n,1}$ appearing in (\ref{pinstantons}) with $\ell=1$. An analysis based on resurgence theory, which can be 
verified with Riemann--Hilbert techniques, leads to the formula \cite{gikm}
\be
u_{n,1}  \sim_n a^{-n+1/2} {S_1\over 2 \pi\ri}  \Gamma\bigl(n-1/2 \bigr)\Bigl\{ 2u_{0,2} + (-1)^n \mu_{0,2}  +  \sum_{l=1}^{\infty} {(2 u_{l,2} +(-1)^{n+l}\mu_{l,2}) a^{l} \over \prod_{m=1}^{l} 
(n-1/2 -m)} \biggr\} 
\ee
where $u_{n,2}$ are the coefficients of the two-instanton expansion ($\ell=2$) in (\ref{pinstantons}), and $\mu_{n,2}$ are the coefficients of the function
\be
u_{1|1}(z)=z^{-3/4}\sum_{n\ge0} \mu_{n,2} z^{-5n/4}.
\ee
It can be seen, by plugging (\ref{ccts}) in the Painlev\'e I equation, that this function satisfies the linear inhomogeneous ODE
\be
-{1\over 6} u_{1|1}'' + 2u_0 u_{1|1}+ 2 u_1 u_{0|1}=0.
\ee
There are similar, but more complicated, formulae for the asymptotic behavior of the coefficients $u_{n,\ell}$ for arbitrary $\ell$, see \cite{gikm}. They all involve the trans-series 
solutions $u_{\ell|1}(z)$. 

 The instanton amplitude $u_{\ell+1}(z)$ can be obtained from the solution $u_{\ell}(z)$ by
 tunneling one extra eigenvalue to the unstable saddle, while the amplitude $u_{\ell-1}(z)$ can be obtained 
from $u_{\ell}(z)$ by tunneling one eigenvalue back to the stable saddle. 
The amplitude $u_{\ell|1}(z)$ does not seem to have, however, an eigenvalue interpretation of this type. The different 
non-perturbative sectors governing the asymptotics of the $\ell$-th instanton solution are depicted in \figref{painlevefig}.

\begin{figure}[!ht]
\leavevmode
\begin{center}
\epsfysize=10cm
\epsfbox{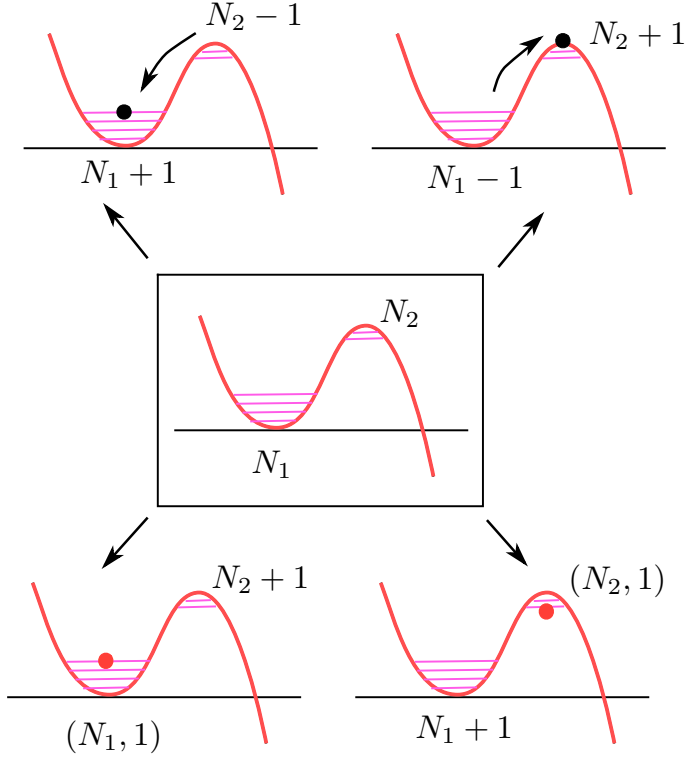}
\end{center}
\caption{In the two-cut case, given a background with perturbative amplitudes $F_g(t_1, t_2)$, there are 
two instantons which are obtained by eigenvalue tunneling, and two 
generalized instanton amplitudes represented by $(N_1, 1)$, $(N_2, 1)$.}
\label{twocutins}
\end{figure}

We can now come back to the original problem of determining the large order behavior of $F_g(S_1, S_2)$, 
corresponding to a two-cut model with $N_1$, $N_2$ eigenvalues around the two 
saddles. There are two instanton configurations which are obtained by eigenvalue tunneling, with fillings $N_1\pm1$, $N_2\mp1$. It can be seen that 
the subleading terms in the asymptotics are not reproduced by using just these two configurations. This is not surprising, in view of the result for the instantons of Painlev\'e I. The above analysis, based on \cite{gikm}, suggests in fact that there should be two other non-perturbative configurations in the two-cut matrix model, 
that we denote by $(N_1, 1)$ and $(N_2,1)$, in analogy with our notation in \figref{painlevefig}. 
These configurations are depicted at the bottom of \figref{twocutins}. The instantons obtained by 
eigenvalue tunneling can be easily calculated from the free energy of the generic two-cut matrix model. However, we do not know how to compute 
the amplitudes involving these new 
configurations, since there is no analogue of the Painlev\'e I equation (or the pre-string equation) for the generic two-cut matrix model. In general, it 
seems that the most general saddle-point of the two-cut matrix model should be labeled by two pairs of integers, $(N_1, M_1)$, $(N_2, M_2)$, associated 
to the two critical points of the cubic potential.

\sectiono{Conclusions and further directions}

In this paper, building on \cite{hk}, we have shown that the direct integration of the holomorphic anomaly equations provides a powerful tool to 
calculate the $1/N$ expansion of multi-cut matrix models. We have seen that, in some circumstances, we can easily fix the holomorphic ambiguity and 
obtain explicit expressions for the genus $g$ amplitudes. In general, we expect the anomaly equation to be integrable, in the sense that 
the gap conditions completely fix the holomorphic ambiguity. In the case of the two-cut cubic matrix model in the slice $S_1=-S_2$, 
we can use this method to determine the amplitudes to very high genus. 

These high genus results have allowed us to obtain quantitative evidence for the connection between large order behavior and eigenvalue tunneling in a multi-cut matrix model. However, our results indicate that the detailed large genus asymptotics of the amplitudes cannot be understood just by considering the non-perturbative sectors associated with eigenvalue tunneling. Indeed, in a similar asymptotic problem analyzed in \cite{gikm}, it was necessary to include new non-perturbative sectors. 
It is only natural to suggest that a correct understanding of the asymptotic properties, in the multi-cut case, requires also the inclusion of new non-perturbative sectors.  
In the one-cut case and its double-scaling limit, the amplitudes in these new sectors can be 
obtained algebraically, as trans-series solutions to the pre-string equation and the Painlev\'e I equation, respectively. In the multi-cut case 
there is no analogue of these equations, and therefore the corresponding generalized amplitudes can not be computed with our present tools.

One obvious question is then the following: what is the interpretation of these new non-perturbative sectors in terms of matrix models or topological strings? We will give now some hints which might help in answering this question. Let us first discuss the trans-series solutions $u_{n|m}(z)$ appearing in (\ref{ccts}). It turns out that $u_{0|\ell}(z)$ can be obtained from $u_{\ell|0}(z)$, the standard instanton amplitude, by changing the sign 
\be
z^{5/4}\rightarrow -z^{5/4}. 
\ee
This corresponds to changing the sign of the string coupling constant $g_s \rightarrow -g_s$. If we think about the 
$u_{\ell}(z)$ as describing a set of $\ell$ D-branes, then 
the natural interpretation of $u_{0|\ell}(z)$ is as a set of $\ell$ {\it anti-D-branes}. Indeed, it has been argued that anti-D-branes are obtained from 
D-branes in topological string 
theory just by changing the sign of the string coupling constant \cite{supervafa}. More generally, these should be the ghost D-branes introduced in \cite{ot}, which 
reduce to anti-D-branes in the topological string context. It is then natural to interpret the generalized instanton amplitude $u_{n|m}(z)$ as representing a
state of $n$ D-branes and $m$ anti-D-branes at the unstable saddle, in the background of $N-n+m$ D-branes in the stable saddle. If this interpretation is correct, the generalized amplitudes in the multi-cut matrix model, which we labeled by two pairs of integers $(N_1, M_1)$, $(N_2, M_2)$, should correspond to a saddle 
where there are $N_i$ branes and $M_i$ anti-D-branes at the $i$-th critical point, $i=1,2$. 

One problem with this interpretation is that, as argued in \cite{supervafa,dv}, such a configuration is described in principle by a quiver or supergroup matrix model. If this 
is the case, the non-perturbative configuration characterized by $(N_i, M_i)$, $i=1,2$, would be equivalent to a configuration with 
only branes or only antibranes at the critical points. More precisely, we would get $|N_i-M_i|$ branes or $|N_i-M_i|$ anti-branes depending on the sign 
of $N_i-M_i$. Since explicit calculations show that the amplitude $u_{n|m}(z)$ is not equal to the 
amplitude $u_{n-m|0}(z)$ \cite{gikm}, the interpretation in terms of brane/anti-brane systems might not be completely appropriate. 

We believe that the appearance of these new sectors indicates that we do not 
fully understand the non-perturbative structure of matrix models and of two-dimensional gravity. Therefore, it would be very important 
to clarify their meaning and to compute their amplitudes in the multi-cut case.

%%%%%%%%%%%%%%%%%%%%%%%%%%%%%%%%%%%%%%%%%%%%%%%%%%%%%%%%%%%%%%%%%%%%%%%%%%%%
\section*{Acknowledgments}

We would like to thank Babak Haghighat, Sara Pasquetti, Ricardo Schiappa and Piotr Su{\l}kowski for useful conversations. 
M.M. would like to specially thank Marlene Weiss for collaboration on this topic in 2008, and Pavel Putrov for discussions and collaboration 
on non-perturbative aspects of matrix models. 

The work of M.M. is supported in part by the Fonds National Suisse. The work of M.R.~is supported by the German Excellence 
Initiative via the graduate school BCGS.

%%%%%%%%%%%%%%%%%%%%%%%%%%%%%%%%%%%%%%%%%%%%%%%%%%%%%%%%%%%%%%%%%%%%%%%%%%%%

\appendix

\sectiono{Data of the two-cut example}\label{AppData}
In the following we collect the necessary data for our two-cut cubic model of the main body text. We restrict ourselves to the points in moduli space which are relevant for our discussion. For further background on e.g.~the monodromy around several divisors in moduli space we refer the reader to \cite{hk}.
\subsection{Large Radius}
$C_1\cap C_2=\{z_1=0\}\cap\{z_2=0\}$:\\
The Picard-Fuchs operators governing the periods of the cubic matrix model are given by
\be\label{eq:PFop}
\begin{split}
\CL_1 =& (3-2z_1-6z_2)\p_1-2z_1(1-2z_1-6z_2)\p_1^2+(1-10z_1+12z_1^2+4z_1z_2)\p_1\p_2 \\
&+(3-6z_1-2z_2)\p_2 + (1-10z_2+4z_1z_2+12z_2^2)\p_1\p_2-2z_2(1-6z_1-2z_2)\p_2^2, \\
\CL_2 =& -3(1-12z_1+18z_1^2+14z_1z_2)+(-3z_2(1-3z_2+2z_2^2) \\
&+z_1(7+46z_1^2-18z_2+26z_2^2+z_1(-39+62z_2)))\p_1 \\
&+(-1+2z_1+2z_2)(-2z_1(1+5z_1^2-2z_1z_2-3z_2^2-4(z_1+z_2))\p_1^2 \\
&+(z_1+z_2)(1-8z_1+6z_1^2-6z_1z_2)\p_1\p_2) \\
&-3(1-12z_2+14z_1z_2+18z_2^2)+(-3z_1(1-3z_1+2z_1^2) \\
&+z_2(7-18z_1+26z_1^2+(-39+62z_1)z_2+46z_2^2))\p_2 \\
&+(-1+2z_1+2z_2)((z_1+z_2)(1-8z_2-6z_1z_2+6z_2^2)\p_1\p_2 \\
&-2z_2(1-3z_1^2-2z_1z_2+5z_2^2-4(z_1+z_2))\p_2^2).
\end{split}
\ee
Its discriminant can be determined to be
\be
\text{disc}=z_1z_2 I^2 J=z_1z_2(1-2(z_1+z_2))(1-6z_1-6z_2+9z_1^2+14z_1z_2+9z_2^2),
\ee
and its solutions around $z_i=0$, $i=1,2$, are given by the following expansions
\be
\begin{split}\label{periodsapp}
S_1&=\frac{z_1}{4}-\frac{1}{8}z_1(2z_1+3z_2)+\dots \\
S_2&=-\frac{z_1}{4}+\frac{1}{8}z_2(3z_1+2z_2)+\dots \\
\Pi_1&=S_1\log\left(\frac{z_1}{4}\right)+\frac{1}{12}-\frac{z_1}{4}-\frac{1}{16}(2z_1^2-10z_1z_2-5z_2^2)+\dots\\
\Pi_2&=S_2\log\left(-\frac{z_2}{4}\right)-\frac{1}{12}+\frac{z_2}{4}-\frac{1}{16}(5z_1^2+10z_1z_2-2z_2^2)+\dots.
\end{split}
\ee
The Yukawa couplings are given by
\be\label{eq:YukFull}
\begin{split}
C_{z_1z_1z_1}&=\frac{1-6z_1+9z_1^2-5z_2+9z_1z_2+6z_2^2}{16z_1I^2} \\
C_{z_1z_1z_2}&=\frac{1-3z_1-5z_2}{16I^2} \\
C_{z_1z_2z_2}&=\frac{1-5z_1-3z_2}{16I^2} \\
C_{z_2z_2z_2}&=\frac{1-5z_1+6z_1^2-6z_2+9z_1z_2+9z_2^2}{16z_2I^2},
\end{split}
\ee
where all other combinations follow by symmetry. The genus one free energy can be written as
\be\label{eq:F1Full}
F_1=-\frac{1}{2}\log\left(\det(G_{i\bar\jmath})\right)-\frac{1}{12}\log(z_1z_2)-\frac{1}{2}\log I + \frac{1}{3}\log J.
\ee
It is convenient to introduce new variables $\tilde{z}_i$, $i=1,2$, by
\be
\tilde{z}_1=z_1+z_2,\quad \tilde{z}_2=\frac{1}{4}(z_1-z_2)\sqrt{1-2(z_1+z_2)},
\ee
as well as coordinates $\tilde{t}_{i}$, $i=1,2$, on the mirror by
\be
\tilde{t}_{1}=s=\frac{1}{2}(S_1-S_2),\quad \tilde{t}_{2}=t=S_1+S_2,
\ee
such that the mirror map becomes as simple as possible. E.g.~we have that
\be
\tilde{z}_2=\tilde{t}_{2}.
\ee
This implies that some of the Christoffel symbols vanish:
\be
\Gamma^{\tilde{z}_2}_{\tilde{z}_i \tilde{z}_j}=0,\quad\text{ for }\quad i=1,2.
\ee
There are only four non-vanishing ambiguities $\tilde f_{ij}^k$ of equation (\ref{PropEq}), that are given by
\be
\begin{split}
\tilde f_{\tilde{z}_1\tilde{z}_1}^{\tilde{z}_1}&=-\frac{5-28\tilde{z}_1+52\tilde{z}_1^2-32\tilde{z}_1^3-112\tilde{z}_2^2}{2(1-2\tilde{z}_1)(1-8\tilde{z}_1+20\tilde{z}_1^2-16\tilde{z}_1^3+16\tilde{z}_2^2)},\\
\tilde f_{\tilde{z}_1\tilde{z}_2}^{\tilde{z}_1}&=\frac{24\tilde{z}_2}{1-8\tilde{z}_1+20\tilde{z}_1^2-16\tilde{z}_1^3+16\tilde{z}_2^2}, \\
\tilde f_{\tilde{z}_2\tilde{z}_2}^{\tilde{z}_1}&=\frac{8-16\tilde{z}_1}{1-8\tilde{z}_1+20\tilde{z}_1^2-16\tilde{z}_1^3+16\tilde{z}_2^2}.
\end{split}
\ee
This results in a propagator that has one non-vanishing component in $\tilde{z}$-coordinates, i.e.
\be
S^{\tilde{z}_1\tilde{z}_1}=4\tilde{z}_1^2-64\tilde{z}_2^2+44\tilde{z}_1^3-832\tilde{z}_1\tilde{z}_2^2+\dots,\qquad S^{\tilde{z}_1\tilde{z}_2}=S^{\tilde{z}_2\tilde{z}_1}=S^{\tilde{z}_2\tilde{z}_2}=0.
\ee
The covariant derivative closes on this propagator when one fixes yet another ambiguity $f^{ij}_k$, cf.~eq.~(\ref{DS}). The only relevant, non-vanishing component is given by
\be
f^{\tilde{z}_1\tilde{z}_1}_{\tilde{z}_1}=\frac{8(1-2\tilde{z}_1)^3(\tilde{z}_1-4\tilde{z}_1^2+4\tilde{z}_1^3-64\tilde{z}_2^2+144\tilde{z}_1\tilde{z}_2^2)}{(1-8\tilde{z}_1+20\tilde{z}_1^2-16\tilde{z}_1^3+16\tilde{z}_2^2)^3}.
\ee

\subsection{Conifold}
Conifold $J=\{1-6z_1-6z_2+9z_1^2+14z_1z_2+9z_2^2=0\}$:\\
We consider the point $(z_1,z_2)=(\frac{1}{8},\frac{1}{8})\in J$. Convenient coordinates are given by
\be
z_{c,1}=\frac{1}{\sqrt{2}}(z_1-z_2),\qquad z_{c,2}=1-4(z_1+z_2).
\ee
$z_{c,1}$ parametrizes the tangential direction to the conifold divisor, whereas $z_{c,2}$ the normal one. Transforming the Picard-Fuchs system to these new coordinates the polynomial solutions are given by
\be
\begin{split}
\omega_1&=z_{c,1}\sqrt{1+z_{c,2}}=z_{c,1}+\frac{1}{2}z_{c,1}z_{c,2}-\frac{1}{8}z_{c,1}z_{c,2}^2+{\cal O}(z_c^4),\\
\omega_2&=z_{c,2}^2+8 z_{c,1}^2 z_{c,2} + {\cal O}(z_c^4).
\end{split}
\ee
We choose as flat coordinates 
\be
t_{c,i}=\omega_i,\quad i=1,2.
\ee
By Inverting the above relations it is easy to calculate the holomorphic limit of the metric and the Christoffel symbols in $z_c$ coordinates. Transforming the Yukawa couplings $C_{ijk}$ as well as the ambiguities $\tilde f_{ij}^k$ yields the propagator at the conifold point. This allows now to expand the free energies $F_g$ in the holomorphic limit at the conifold point.

\sectiono{The one-cut solution}
\label{sec:onecut}
A special case of the multi-cut matrix model occurs when all the partial 't Hooft parameters are zero except for one, $S_2=\cdots=S_n=0$. This is 
called the {\it one-cut} matrix model. A powerful, recursive solution of the one-cut matrix model at all genera has been known for a long time \cite{bessis,biz}, 
and it is based on the technique of orthogonal polynomials. In this appendix we list some ingredients of the one-cut solution. 

We first recall that the orthogonal polynomials $p_n(\lambda)$ for the potential $V(\lambda)$ are defined by
\be
\int {\rd\lambda\over 2\pi}\, \re^{-{V(\lambda)\over g_s}}\, p_n(\lambda) p_m(\lambda)= h_n\, \delta_{nm}, \qquad n>0,
\ee
\noindent
where $p_n$ are normalized by requiring that $p_n\sim \lambda^n+\cdots$. It is well known (see for example \cite{dfgzj}) that 
the partition function of the matrix model can be expressed as
\be
Z_N=\prod_{i=0}^{N-1}h_i=h_0^N\, \prod_{i=1}^N r_i^{N-i}.
\ee
\noindent
The coefficients 
\be
r_n={h_n\over h_{n-1}}
\ee
\noindent
satisfy recursion relations depending on the shape of the potential. They also obviously satisfy
\be\label{rZ}
r_n={Z_{n-1}Z_{n+1}\over Z_n^2}.
\ee 
\noindent
In the limit $N\rightarrow \infty$, $n/N$ becomes a continuous variable that we will denote by $z$, and $r_n$ is promoted to a function, $R(z,g_s)$. The perturbative $g_s$ 
expansion is obtained by writing
\be
R(z,g_s)=\sum_{\ell=0}^{\infty} R_{\ell}(z) g_s^{2\ell}.
\ee

We now consider the cubic matrix model with potential
\be
{1\over g_s} W(z)={1\over g_s} \left( -z+{z^3\over 3}\right).
\ee
The resulting matrix model is equivalent to the one we considered in the bulk of the paper, up to a linear transformation $z\rightarrow az +b$, but 
leads to a simple recursion relation for the coefficients $r_n$ (see for example \cite{dfgzj})
\be\label{oprecursion}
r_n\left(\sqrt{1-r_n-r_{n+1}}+\sqrt{1-r_n-r_{n-1}}\right)=g_s n.
\ee
\noindent
The continuum limit of equation \eqref{oprecursion} above is
\be\label{Rdifference}
R(t,g_s)\left(\sqrt{1-R(t,g_s)-R(t+g_s,g_s)}+\sqrt{1-R(t,g_s)-R(t-g_s,g_s)}\right)=t.
\ee
\noindent
At lowest order in $\ell$ and $g_s$ we find
\be
2R_0(t)\sqrt{1-2R_0(t)}=t.
\ee
\noindent
It turns out to be convenient to express everything in terms of 
\be
U(t,g_s) = {\sqrt{1-2 R(t,g_s)}}
\ee
This has a $g_s$ expansion of the form
\be
U(t,g_s) =\sum_{\ell\ge 0} U_{\ell}(t) g_s^{2g}. 
\ee
Notice that 
\be
u\equiv U_0(t) 
\ee
satisfies 
\be
u(1-u^2) =t,
\ee
and one chooses the solution with expansion
\be
u=1-{t\over 2} -{3 t^2 \over 2} -\cdots
\ee
We also have the inverse relation
\be
R(t,g_s)=\frac{1}{2}(1 -U^2(t,g_s)).
\ee
This variable eliminates square roots. The recursion becomes
\be
\label{udifference}
U (1-U^2)\left\{ \sqrt{1+\Bigl({U^+\over U}\Bigr)^2 }+\sqrt{1+\Bigl({U^-\over U}\Bigr)^2 } \right\}=2{\sqrt{2}} t
\ee
where 
\be
U^{\pm}=U(t\pm g_s,g_s). 
\ee
Expanding the equation \eqref{udifference} in power series of $g_s$, and solving it recursively, one can compute in this way the coefficients $U_{n}(u)$, and from them derive the 
coefficients $R_{\ell}$. The first few read, when expressed in terms of 
\be
r={1-u^2\over 2}
\ee
as
\be
R_1(t) = -\frac{ (9 r-5)}{32 (1-3 r)^4}, \qquad R_2(t) = -\frac{3  \left(162 r^3+1017 r^2-1316 r+385\right)}{2048 (3 r-1)^9}.
\ee

Once $R(z,g_s)$ are calculated, one can easily calculate the total free energy 
\be
F(t,g_s)=\sum_{g\ge 0} g_s^{2g-2}F_g(t). 
\ee
Let us define
\be
\Xi(z,g_s) = {R(z,g_s) \over z}
\ee
One then obtains \cite{bessis,biz}:
\be
\label{ofgex}
\ba
g_s^2 F &= \int_0^t \rd z\,  (t-z) \log \Xi (z)+  \sum_{p=1}^{\infty} g_s^{2p}\, \, {B_{2p} \over (2p)!}\, \frac{\rd ^{2p-1}}{\rd z^{2p-1}} \biggl[ \left( t-z \right) \log \Xi(z,g_s) \biggr] \bigg|_{z=0}^{z=t} \\
&+ {t g_s \over 2 } \biggl[ 2 \log {h_0 \over h_0^{\rm G}} - \log \Xi (0,g_s)\biggr].
\ea
\ee
Here, $h_0/h_G$ is the integral
\be
{h_0 \over h_G}={1\over {\sqrt{2\pi}}} \int_{-\infty}^{\infty} \re^{-x^2/g_s -x^3/(3 g_s)} \rd x
\ee
understood as a formal power series in $g_s$. This can be explicitly calculated
\be
{h_0 \over h_G}={1\over \pi^{1/2}} \sum_{k\ge 0} {\Gamma\Bigl( {1\over 2}  + 3k\Bigr) \over 3^{2k} (2k)!} g_s^k.
\ee
Notice that, as pointed out in \cite{biz} in the case of a quartic potential, the $g_s$ expansion of 
\be
2 \log {h_0 \over h_0^{\rm G}} - \log \Xi (0,g_s)
\ee
contains only odd powers of $g_s$, as required in order to have an expansion of the free energy in terms of even powers of $g_s$. 
Notice that the explicit solution for the $F_g$ in the one cut case provides an additional boundary condition for the holomorphic anomaly of matrix models . 

%%%%%%%%%%%%%%%%%%%%%%%%%%%%%%%%%%%%%%%%%%%%%%%%%%%%%%%%%%%%%%%%%%%%%%%%%%%%%%%%
\sectiono{Modular forms and elliptic integrals}\label{AppModEll}
We follow the conventions in \cite{bf}. The complete elliptic integral of the first kind is defined as
\be
K(k)=\int_0^1 \frac{\rd t}{\sqrt{(1-t^2)(1-k^2t^2)}}.
\ee
The parameter $k$ is called the elliptic modulus. Further one defines the complementary modulus as ${k'}^2=1-k^2$. The complete elliptic integral of the second kind is defined as
\be
E(k)=\int_0^1 \rd t\sqrt{\frac{1-k^2t^2}{1-t^2}}.
\ee
The complete elliptic integrals of the first and second kind are related to each other by derivation,
\be
\frac{\rd K}{\rd k} = \frac{E(k)-k'^2 K(k)}{kk'^2},\qquad \frac{\rd E}{\rd k} = \frac{E(k)-K(k)}{k}.
\ee
Useful transformation formulae are
\be
\begin{split}
K\left(\frac{1-k'}{1+k'}\right)&=\frac{1+k'}{2}K(k),\\
E\left(\frac{1-k'}{1+k'}\right)&=\frac{1}{1+k'}(E(k)+k'K(k)),\\
K\left(\frac{2\sqrt{k}}{1+k}\right)&=(1+k)K(k),
\end{split}
\ee
as well as the Legendre relation
\be
E(k)K(k')+E(k')K(k)-K(k)K(k')=\frac{\pi}{2}.
\ee
Consider an elliptic geometry of the form
\be\label{geometryapp}
y^2=\prod_{i=1}^4 (x-x_i),
\ee
where $x_1<x_2<x_3<x_4$ are the branch cuts. Define the half-period ratio of the elliptic geometry, $\tau$, and the elliptic nome as $q=e^{\ri \pi \tau}$. It can be shown that
\be
\tau=\ri\frac{K(k')}{K(k)},
\ee
and moreover that
\be
K(k)=\frac{\pi}{2}\vartheta_3^2,\quad k^2=\frac{\vartheta_2^4}{\vartheta_3^4},\quad k'^2=\frac{\vartheta_4^4}{\vartheta_3^4}.
\ee
Here $\vartheta_i$ are the Jacobi theta-functions defined by
\be
\vartheta_2=\sum_{n\in\IZ}q^{\frac{1}{2}(n+\frac{1}{2})^2},\quad \vartheta_3=\sum_{n\in\IZ}q^{\frac{1}{2}n^2},\quad \vartheta_4=\sum_{n\in\IZ}(-1)^n q^{\frac{1}{2}n^2}.
\ee

The Thomae formula relates the branch cuts of an elliptic curve to theta-functions \cite{fay}. For the geometry consider above (\ref{geometryapp}) we obtain
\be
\begin{split}
\vartheta_2^4(\tau)&=-\CK^2(x_1-x_2)(x_3-x_4) \\
\vartheta_3^4(\tau)&=-\CK^2(x_1-x_4)(x_2-x_3) \\
\vartheta_4^4(\tau)&=-\CK^2(x_1-x_3)(x_2-x_4),
\end{split}
\ee
and thus
\be
\eta^{24}(\tau)=\frac{\CK^{12}}{256}\, \prod_{i\, <\, j}(x_i-x_j)^2,
\ee
where $\CK$ and $\CK'$ are given in (\ref{Kdef}).

It is convenient to introduce
\be
b=\vartheta_2^4,\quad c=\vartheta_3^4, \quad d=\vartheta_4^4,
\ee
where either two of them span the ring of $\Gamma(2)$ modular forms. Here the congruence subgroup $\Gamma(2)\subset\text{SL}(2,\IZ)$ is defined by
\be
\Gamma(2)=\{\gamma\in\text{SL}(2,\IZ)\,|\,\gamma\equiv\mathds{1}\text{ mod }2\}.
\ee

%%%%%%%%%%%%%%%%%%%%%%%%%%%%%%%%%%%%%%%%%%%%%%%%%%%%%%%%%%%%%%%%%%%%%%%%%%%%%%%%

\end{document}